\documentclass[11pt,a4paper]{article}

\usepackage[colorlinks=true, linkcolor=black!50!blue, urlcolor=blue, citecolor=blue, anchorcolor=blue]{hyperref}
\usepackage[font=small,labelfont=bf,margin=0mm,labelsep=period,tableposition=top]{caption}
\usepackage[a4paper,top=3cm,bottom=2.5cm,left=2.5cm,right=2.5cm,bindingoffset=0mm]{geometry}

\usepackage{graphicx,placeins}
\usepackage{float}
\usepackage{afterpage}
\usepackage{epsfig,cite}
\usepackage{amssymb}
\usepackage{amsmath}
\usepackage{dsfont}
\usepackage{multirow}
\usepackage{url}
\usepackage{xcolor}
\usepackage{float}
\usepackage{afterpage}

\usepackage{url}
\usepackage{hyperref}

\usepackage{booktabs}

\usepackage{tikz}
\usetikzlibrary{arrows}
\usepackage{tikz-3dplot}
\usepackage{enumitem}
\usepackage{hyperref}
\usepackage{cite}

\def\smallfrac#1#2{\hbox{$\frac{#1}{#2}$}}
\def\half{\smallfrac{1}{2}}

\newcommand{\be}{\begin{equation}}
\newcommand{\ee}{\end{equation}}
\newcommand{\bea}{\begin{eqnarray}}
\newcommand{\eea}{\end{eqnarray}}
\newcommand{\bi}{\begin{itemize}}
\newcommand{\ei}{\end{itemize}}
\newcommand{\ben}{\begin{enumerate}}
\newcommand{\een}{\end{enumerate}}

\def\frac#1#2{{{#1}\over {#2}}}
\def\gsim{\mathrel{\rlap{\lower4pt\hbox{\hskip1pt$\sim$}}
    \raise1pt\hbox{$>$}}}         
\def\lsim{\mathrel{\rlap{\lower4pt\hbox{\hskip1pt$\sim$}}
    \raise1pt\hbox{$<$}}}         
\def\nn{\nonumber}

\newcommand{\draft}[1]{}

\def\beq{\begin{equation}}
\def\eeq{\end{equation}}

\def\lapprox{\lower .7ex\hbox{$\;\stackrel{\textstyle <}{\sim}\;$}}
\def\gapprox{\lower .7ex\hbox{$\;\stackrel{\textstyle >}{\sim}\;$}}
\def\half{\smallfrac{1}{2}}

\def\Var{{\rm Var}}
\def\Cov{{\rm Cov}}
\def\betatil{\widetilde{\beta}}
\def\lambdatil{\widetilde{\lambda}}
\def\lambdabar{\overline{\lambda}}
\def\Ttil{\widetilde{T}}

\def\Stil{\widetilde{S}}
\def\Xtil{\widetilde{X}}
\def\Ptil{\widetilde{P}}

\def\Shat{\widehat{S}}
\def\Xhat{\widehat{X}}

\def\Zbar{\overline{Z}}

\def\Tdot{\dot{T}}
\def\Ttildot{\skew6\dot\Ttil}

\numberwithin{equation}{section}
\numberwithin{figure}{section}
\numberwithin{table}{section}

\usepackage{tabularx}
\newcolumntype{C}[1]{>{\centering\arraybackslash}p{#1}}

\begin{document}
\newgeometry{top=1.5cm,bottom=1.5cm,left=2.5cm,right=2.5cm,bindingoffset=0mm}
\vspace{2.0cm}
\begin{flushright}
Edinburgh 2019/17\\
\end{flushright}
\vspace{1.3cm}

\begin{center}
  {\Large \bf Correlation of Theoretical Uncertainties in PDF Fits and Theoretical Uncertainties in Predictions}
\vspace{2.1cm}

 {\small
Richard D. Ball and 
Rosalyn L. Pearson
}\\

 \vspace{0.7cm}
 
       {\it \small 
         ~The Higgs Centre for Theoretical Physics, University of Edinburgh,\\
  JCMB, KB, Mayfield Rd, Edinburgh EH9 3FD, Scotland\\[0.1cm]
}

\vspace{3.0cm}

{\bf \large Abstract}

\end{center}
We show how to account for correlations between theoretical uncertainties incorporated in parton distribution function (PDF) fits, and the theoretical uncertainties in the predictions made using these PDFs. We demonstrate by explicit calculations, both analytical and numerical, that these correlations can lead to corrections to the central values of the predictions, and reductions in both the PDF uncertainties and the theoretical uncertainties in the prediction. We illustrate our results with predictions for top production rapidity distributions and the Higgs total cross-section at the LHC, using the NLO NNPDF3.1 PDF set which incorporates missing higher order uncertainties. We conclude that the inclusion of correlations can increase both the accuracy and precision of predictions involving PDFs, particularly for processes with data already included in the PDF fit.

\clearpage


Parton distribution functions (PDFs) play a crucial role in the prediction of Standard Model observables at the Large Hadron Collider (LHC) and beyond~\cite{Forte:2020yip, Ethier:2020way}. PDF uncertainties are already a limiting factor in these predictions~\cite{Azzi:2019yne}. This means that a precise and accurate knowledge of PDFs is essential for full exploitation of the physics at LHC. Besides the familiar experimental uncertainties in the data that are input to a global PDF fit, it is becoming increasingly necessary to also consider theoretical uncertainties: missing higher order uncertainties (MHOUs) due to the truncation of perturbative expansions in the hard cross-sections and parton evolution, uncertainties due to the use of nuclear targets, uncertainties due to missing higher twist, uncertainties due to external parameters such as quark masses, showering and hadronization uncertainties in final state Monte Carlos, and so on. In the past the technique used to estimate the effect of theoretical uncertainties was to compare the PDFs produced with and without various theoretical corrections, rather than to incorporate the uncertainties into the fit itself. However the limitations of such a procedure are clear.

Recently a new approach to estimating the impact of theoretical uncertainties on global PDF fits has been developed, through the construction of a `theory covariance matrix' \cite{Ball:2018lag}, analogous to the experimental covariance matrix used in global PDF fits. By adding the theory covariance matrix to the experimental covariance matrix as an additional source of uncertainty, theoretical uncertainties can be incorporated directly in the fit, where they impact not only the overall level of PDF uncertainty, but also the relative weight of different datasets. This novel approach has so far been applied to uncertainties associated with nuclear effects \cite{Ball:2018twp,Ball:2020xqw}, and to the estimation of MHOUs by scale variation \cite{Pearson:2018tim, AbdulKhalek:2019bux, AbdulKhalek:2019ihb} (though other methods~\cite{Cacciari:2011ze, Bagnaschi:2014wea, Bonvini:2020xeo} of estimating MHOU are under development). Factorization scale variations estimate the MHOUs in parton evolution (correlated across all observables), and renormalization scale variations estimate  the MHOUs in fixed order calculations of process-dependent hard cross-section (correlated only across a given class of processes). The first global PDFs including MHOUs estimated in this way were presented in \cite{AbdulKhalek:2019ihb}. 

When making predictions for hadronic observables there are again two sources of MHOU: uncertainties in the PDF evolution, which can also be estimated by factorization scale variation, and uncertainties in the hard cross-section, again estimated by renormalization scale variation. These arise on top of the MHOUs incorporated in the determination of the PDFs themselves, manifested as a part of the PDF uncertainty. So there will be correlations between the MHOUs in the PDFs and the additional MHOUs in the predictions. The PDFs themselves contain a wealth of data from a wide range of processes. When making a prediction for one of these processes (for example in a new kinematic regime), there will inevitably be correlations between the renormalization scale variation in the PDF determination, and that in the prediction. Even if the process is a new one, for example Higgs production, correlations due to factorization scale variation will still be present: all processes dependent on PDFs have a MHOU due to the need to evolve the PDFs to the scale of the process. 

The potential importance of these correlations can be exposed by considering a simple situation in which a PDF is determined from data at a given scale on a single observable (such as a nonsinglet structure function), and used to make predictions of another observable at the same scale \cite{Harland-Lang:2018bxd}. In this special case the PDF can be eliminated altogether, and the predicted observable determined directly from the measured observable. Including the MHOU twice (once in the calculation of the measured observable, then again in making the predicted observable), whilst ignoring their possible correlation, would amount to ``double counting", and thus an overestimate of the uncertainty. Of course in a global PDF fit, involving many different processes, this particular formulation is no longer applicable. However the fact remains that there will still be some residual correlation between the MHOU in the PDF determination and the MHOU in the prediction, and to neglect it may lead to an overestimation of the MHOU. 

In Ref.~\cite{AbdulKhalek:2019ihb}, it was seen that in a realistic NLO global fit, the effect of MHOU on the PDF uncertainty is 
quite small, the main consequence being a rebalancing of the impact of different datasets depending on their relative MHOU. The MHOU in the PDF is consequently rather smaller than the MHOU in the prediction, and if they are combined in quadrature, the effect of missing correlations will likely also be small. It was further argued that combining the PDF uncertainties and theoretical uncertainties in quadrature is conservative, since it can only lead to an overestimate of uncertainties, and thus better than neglecting MHOUs altogether. Nevertheless, since the correlation between the uncertainties has not been computed explicitly, there remains the intriguing possibility that in some circumstances including the correlation may yield more precise, and perhaps even more accurate, predictions. 

In this paper we show explicitly how to propagate theoretical uncertainties in the determination of global PDFs into the predictions made using these PDFs, taking account of all correlations between theoretical uncertainties. As this is a complicated problem, we proceed step by step. In Sec.~1 we show how a single source of theoretical uncertainty can be reformulated in terms of a nuisance parameter, which holds the key to the propagation of uncertainties. We explore the interplay between the theoretical uncertainties and the experimental uncertainties in the data in two idealised contexts: the first in which there are no fitted parameters, the second in which there as many parameters as data, so the experimental data can be fitted exactly. This allows us to identify three distinct effects of the correlation of theoretical uncertainties. Then in Sec.~2 we consider a more realistic, but still simplified, situation where we fit the data using a single parameter, still in a theory with a single source of theoretical uncertainty, and find that all three of these correlation effects are still present, and can be computed.  In Sec.~3 these results are extended to multiple theoretical uncertainties, in a  multiparameter fit, and then less trivially to a PDF fit where the PDFs are continuous functions, with a functional uncertainty. Finally in Sec.~4 we present numerical results, made in the context of the NNPDF3.1 NLO global fit with MHOU presented in Ref.~\cite{AbdulKhalek:2019ihb}, and make predictions with MHOU for repetitions of the experiments included in the fit (so-called `autopredictions'), and for genuine predictions (top and Higgs production). We are able to compute all correlations explicitly, calculating corrections to central values and changes in PDF and theoretical uncertainties for both autopredictions and genuine predictions, and confirm that the correlated predictions can be both more accurate and more precise than the conservative prescription. A summary is provided in Sec.~5.

\section{Predictions with Correlated Theoretical Uncertainties}
\label{sec:generic}

We showed in Refs.\cite{Ball:2018lag, Ball:2018twp} that when we fit $N$ experimental data points $D_i$ to theoretical predictions $T_i$, $i = 1,\ldots,N$, then the uncertainties in the theoretical predictions can can be incorporated into the fit simply by adding a theoretical covariance matrix $S_{ij}$ to the experimental covariance matrix $C_{ij}$. The only assumptions made in deriving this result are that all uncertainties, both experimental and theoretical, can be treated as Gaussian, and that the theoretical uncertainties are independent of the experimental data. We used this result in 
Refs.~\cite{Ball:2018twp,Ball:2020xqw} to include nuclear uncertainties in a PDF fit, and in 
Refs.~\cite{AbdulKhalek:2019bux,AbdulKhalek:2019ihb} to incorporate missing higher order uncertainties in global PDF fits.

The result of Sec.~2 of Ref.\cite{Ball:2018twp} may be summarised in terms of the Bayesian probability
\be
\label{eq:PTD}
P(T|D)\propto \exp\left(-\half(T-D)^T(C+S)^{-1}(T-D)\right)
\ee
where for simplicity we adopt a matrix notation: $C$ and $S$ are real symmetric matrices, with $C$ strictly positive definite (thus invertible), and $S$ positive semi-definite. In practice theoretical uncertainties are highly correlated, so $S$ may (and generally will) have zero eigenvalues. We determine $T$ from $D$ by maximizing $P(T|D)$: this is equivalent to minimizing 
\be
\label{eq:chisq}
\chi^2 = (T-D)^T(C+S)^{-1}(T-D)
\ee 
with respect to free parameters characterizing the theoretical prediction.

In this section we will also assume that there is only a single source of fully correlated theoretical uncertainty, so that the theory covariance matrix can be written as 
\be
\label{eq:Sdef}
S = \beta\beta^T
\ee
for some real nonzero vector $\beta$ (i.e. that $S_{ij} = \beta_i\beta_j)$. Then all eigenvalues of $S$ except one are zero. We will develop a nuisance parameter formalism to propagate this theoretical uncertainty (Sec.~\ref{subsec:nuisance1}), and then show how to apply it in two extreme cases: firstly to the situation in which the theory contains no free parameters, and thus where there is no fitting  (Sec.~\ref{subsec:puretheory}), and then to the situation where there as many free parameters as data points, so that we can achieve a perfect fit  (Sec.~\ref{subsec:phenomenology}). 

\subsection{Nuisance Parameters}
\label{subsec:nuisance1}

It will be useful in what follows to introduce a nuisance parameter $\lambda$ for the correlated theoretical uncertainty. Following the notation in Ref.~\cite{Ball:2012wy}, we model the theoretical uncertainty as a fully correlated shift in the theoretical prediction: $T\to T+\lambda\beta$. The nuisance parameter $\lambda$ then gives the size of the shift. For a given shift, with Gaussian experimental uncertainties
\be
\label{eq:PTDl}
P(T|D\lambda)\propto \exp\left(-\half(T+\lambda\beta-D)^TC^{-1}(T+\lambda\beta-D)\right).
\ee
Now using Bayes' Theorem,
\be
\label{eq:bayes}
P(T|D\lambda)P(\lambda|D) = P(\lambda|TD)P(T|D).
\ee
To determine $P(T|D)$ we need to first fix the prior distribution of $\lambda$. Since this is a theoretical uncertainty, it is reasonable to assume that the prior is independent of the data, thus that $P(\lambda|D)=P(\lambda)$. Then, marginalizing Eq.~(\ref{eq:prior}) over $\lambda$,
\be
\label{eq:marginalise1}
P(T|D) = \int\! d\lambda\, P(T|D\lambda)P(\lambda). 
\ee
Taking $P(\lambda)$ as a Gaussian, centred on zero (so the theoretical uncertainty is unbiased), with unit width (fixing the overall normalization of the prior theoretical uncertainty), we choose
\be
\label{eq:prior}
P(\lambda) \propto \exp\left(-\half\lambda^2\right).
\ee

The integration over the nuisance parameter is now Gaussian, 
\be
\label{eq:marginalise_gauss}
P(T|D) \propto\int d\lambda\, \exp\left(-\half[(T+\lambda\beta-D)^TC^{-1}(T+\lambda\beta-D)+\lambda^2]\right)\, , 
\ee
and can be performed in the usual way by completing the square:
\begin{eqnarray}
\label{eq:completesquare}
(T+\lambda\beta-D)^TC^{-1}(T+\lambda\beta-D)+\lambda^2
&=&
 Z^{-1}\left(\lambda
+Z\beta^TC^{-1}(T-D)\right)^2\\ && + (T-D)^TC^{-1}(T-D) - Z(\beta^TC^{-1}(T-D))^2,\nn
\end{eqnarray}
where we have defined
\be
\label{eq:Zdef}
Z = (1+\beta^TC^{-1}\beta)^{-1} = 1-\beta^T(C+S)^{-1}\beta.
\ee
The second expression was obtained by noting that 
\be
\label{eq:Zalgebra}
(1+\beta^T C^{-1}\beta)(1-\beta^T(C+S)^{-1}\beta)=1 
\ee
using Eq.~(\ref{eq:Sdef}).

Since
\be
\label{eq:numa}
(\beta^TC^{-1}(T-D))^2 = (T-D)^TC^{-1}\beta\beta^TC^{-1}(T-D),
\ee
we can combine the two terms on the second line of Eq.~(\ref{eq:completesquare})
to give
\be
\label{eq:chisqtoo}
(T-D)^T(C^{-1}-ZC^{-1}\beta\beta^TC^{-1})(T-D) = (T-D)^T(C+\beta\beta^T)^{-1}(T-D),
\ee
since
\be
\label{eq:inverse}
(C+\beta\beta^T)(C^{-1}-ZC^{-1}\beta\beta^TC^{-1}) =1 + \beta\beta^TC^{-1}- Z\beta\beta^TC^{-1} - Z\beta\beta^TC^{-1}\beta\beta^TC^{-1} = 1,  
\ee
by substituting (from Eq.~(\ref{eq:Zdef})) $\beta^TC^{-1}\beta = Z^{-1}-1$ in the last term. Using the definition Eq.~(\ref{eq:Sdef}), we recognise Eq.~(\ref{eq:chisqtoo}) as the $\chi^2$, Eq.~(\ref{eq:chisq}). Defining
\be
\label{eq:lambdabar}
\overline{\lambda}(T,D) = Z\beta^TC^{-1}(D-T)=\beta^T(C+S)^{-1}(D-T),
\ee
we can thus write Eq.~(\ref{eq:marginalise_gauss}) as
\be
\label{eq:integration}
P(T|D)\propto\int d\lambda\, \exp\left(-\half Z^{-1}(\lambda-\overline{\lambda})^2 - \half\chi^2\right) \propto \exp(-\half\chi^2),
\ee
since the integration over $\lambda$ yields a factor $(2\pi Z)^{1/2}$ which 
is independent of $T$ and $D$.

The nuisance parameter formalism thus produces our original result Eq.~(\ref{eq:PTD}). However it is useful because it allows us to also determine the posterior distribution of the nuisance parameter: inverting Eq.~(\ref{eq:bayes}) 
\be
\label{eq:posterior}
P(\lambda|TD)\propto \exp\left(-\half Z^{-1}(\lambda-\overline{\lambda}(T,D))^2\right),
\ee
so once we use the information on $T$ and $D$, the prior distribution 
Eq.~(\ref{eq:prior}) is modified. In particular the peak of the distribution is shifted away from zero to $\overline{\lambda}$, while the width is now given by $Z$. It is easy to see from the definition Eq.~(\ref{eq:Zdef}) that
\be
\label{zbounds}
0 < Z < 1,
\ee
so the width of the theoretical uncertainty is generally reduced by the addition of new information. 

\subsection{Predictions and Autopredictions without Fits}
\label{subsec:puretheory}

In order to explore the implications of the formalism described in the previous section we first consider a `pure' theory $T=T_0$, one where there are no unknown parameters to be fitted, but which nevertheless has a theoretical uncertainty. Despite the fact that this theory cannot be fitted to the data, so in general $T_0\neq D$, the data can still inform the theoretical predictions.

Computing expectation values of functions of $\lambda$ using the probability distribution $P(\lambda|T_0 D)$, 
\be
\label{eq:Elampt}
{\rm E}[\lambda] = {\cal N}_\lambda \int d\lambda\;\lambda\; P(\lambda|T_0 D) = \overline{\lambda}(T_0,D),
\ee
the normalization  ${\cal N}_\lambda$ being chosen such that ${\rm E}[1] =1$, while
\be
\label{eq:varlampt}
\Var[\lambda] \equiv {\rm E}[(\lambda -{\rm E}[\lambda])^2] = Z.
\ee
Now in the nuisance parameter formalism, the theoretical predictions are 
\be
\label{eq:Tlambda}
T(\lambda)=T_0 +\lambda\beta.
\ee 
If we made these predictions before comparing to any data, we would use the prior distribution Eq.~(\ref{eq:prior}) for $\lambda$. Since this is a unit gaussian centred on zero, we would then find that ${\rm E}[T(\lambda)] = T_0$, while $\Cov[T(\lambda)] = \beta\beta^T=S$ as expected. 

However if instead we first compare $T$ to the data $D$, then we can instead compute expectation values using $P(\lambda|TD)$. We then find (using Eqs.~(\ref{eq:Elampt},\ref{eq:lambdabar})) 
 \be
\label{eq:ETlampt}
{\rm E}[T(\lambda)] = T_0+\overline{\lambda}(T_0,D)\beta = T_0 + \beta\beta^T(C+S)^{-1}(D-T_0),
\ee
while (using Eq.~(\ref{eq:varlampt})) 
\be
\label{eq:varTlampt}
\Cov[T(\lambda)] \equiv {\rm E}[(T(\lambda) -{\rm E}[T(\lambda)])(T(\lambda) -E[T(\lambda)])^T] =\Var[\lambda]\beta\beta^T =ZS.
\ee
One can consider this as an `autoprediction': first the theory is compared to the data, and then using this information one makes new theoretical predictions for precise repetitions of the same experiments. The original theoretical predictions $T_0$ are then shifted by an amount
\be 
\label{eq:delTpt}
\delta T = -S(C+S)^{-1}(T_0-D)
\ee
and their theoretical uncertainties are reduced by a factor of $\sqrt{Z}$, since the data add new information: the covariance matrix of the autopredictions is
\be
\label{eq:ZS}
ZS = S - S(C+S)^{-1}S = C(C+S)^{-1}S = S(C+S)^{-1}C.
\ee
This is a simple example of Bayesian learning: the theory `learns' from the data, within the constraints imposed by the prior theoretical uncertainty.

It is interesting to compare the experimental $\chi^2$ of the original predictions 
\be
\label{eq:expchisq}
\chi^2_{\rm exp} = (T_0-D)^TC^{-1}(T_0-D)
\ee
to that obtained with the autopredictions
\bea
\chi^2_{\rm auto} &=& (T_0+\delta T-D)^T C^{-1}(T_0+\delta T-D)\nn\\
&=& (T_0-D)^T (C+S)^{-1}C(C+S)^{-1}(T_0-D),
\label{eq:chisqauto}
\eea  
since
\be
\label{eq:TplusdelTminD}
T_0+\delta T-D = C(C+S)^{-1}(T_0-D).
\ee
It is easy to see that $\chi^2_{\rm auto}\leq\chi^2_{\rm exp}$ since $C$ is positive definite (i.e. $C>0$) and $S$ semi-positive definite (i.e. $S\geq 0$), so $2S+SC^{-1}S\geq 0$, whence $(C+S)C^{-1}(C+S)\geq C$, and $(C+S)^{-1}C(C+S)^{-1}\leq C^{-1}$. So the data induced shifts are always such as to improve the quality of the fit to the data, by exploiting the theoretical uncertainty.

To make this more explicit, consider a simple model in which the experimental uncertainties are uncorrelated, and the same for each data point: then we can write 
\be
\label{eq:modelCS}
C = \sigma^2 1,\qquad S = s^2 e e^T,
\ee
with $e$ a unit vector, $e^Te=1$, and $\beta = s e$ so $S=\beta\beta^T$. Thus $\sigma$ is the experimental uncertainty on each data point, and $s/\sqrt{N}$ is the size of the correlated theoretical uncertainty. Then it is easy to see that
\be
\label{eq:modelCplusSinv}
(C+S)^{-1} = \frac{1}{\sigma^2}\left(1-\frac{s^2}{\sigma^2+s^2}e e^T\right),
\ee
and (using Eq.~(\ref{eq:Zdef}))
\be
\label{eq:modelZS} 
Z = (1+s^2/\sigma^2)^{-1}.
\ee 
So the reduction in theoretical uncertainties depends on the ratio $s^2/\sigma^2$: when $s^2\ll\sigma^2$ the influence of the data on the theoretical uncertainty is very small,
while when $s^2\gg\sigma^2$ the size of the theoretical uncertainty is reduced from $s$ to $\sigma$, since in the limit $s^2/\sigma^2\to \infty$ 
\be
\label{eq:modelZ} 
ZS = \frac{\sigma^2 s^2}{\sigma^2+s^2}ee^T \to \sigma^2 ee^T.
\ee 
Note that when the theoretical uncertainties are comparable to the experimental, $s^2/N\sim\sigma^2$, and $Z \sim 1/(N+1)$: if there are a large number of independent data points, the reduction of the theoretical uncertainty can be very substantial.

In this model the shifts in the theoretical predictions is (using Eq.~(\ref{eq:delTpt}))
\be 
\label{eq:modeldelTpt}
\delta T = - \frac{s^2}{\sigma^2+s^2} (e^T(T_0-D)) e,
\ee
as expected in the direction $e$ of the theoretical uncertainty. When $s^2/\sigma^2\to \infty$, $e^T(T_0+\delta T)\to e^TD$, so in this direction the autopredictions coincide precisely with the data. The experimental $\chi^2$ for the autopredictions is  (using Eq.~(\ref{eq:chisqauto})) 
\be
\label{eq:modelchisqauto}
\chi^2_{\rm auto} = (T_0-D)^T \frac{1}{\sigma^2}\Big(1 - \frac{s^2(s^2+2\sigma^2)}{(s^2+\sigma^2)^2}ee^T\Big)(T_0-D),
\ee 
So $N-1$ contributions to the $\chi^2$ orthogonal to $e$ are unchanged, while the contribution along $e$ is reduced by a factor $Z^2$.  The autopredictions will in general have a $\chi^2$ of size $N-1$, rather than the $N$ of the original predictions, as naively expected since the nuisance parameter is effectively fitted.

Of course we can also consider genuine predictions $\Ttil_I$, $I=1,\ldots,\widetilde{N}$, which again have no free parameters, but have a theoretical uncertainty which is correlated to that of the observables $T_i$, $i=1,\ldots, N$ for which we have the data $D_i$.  The theoretical predictions including the theoretical uncertainty may be written
\be
\label{eq:Ttillambda}
\Ttil(\lambdatil) =  \Ttil + \lambdatil\betatil,
\ee
where the vector $\betatil_I$ gives the size and direction  of the theoretical uncertainty in $\Ttil_I$.  Now if the nuisance parameters $\lambdatil$ are independent of the parameters $\lambda$, the theoretical uncertainties in $\Ttil$ are uncorrelated with those in $T$, and the theory covariance matrix for the prediction $\Ttil$ is given by 
\be
\Stil = \betatil\betatil^T,\label{eq:Stildef}
\ee

However if they are correlated, $\lambdatil=\lambda$, and then taking advantage of the data $D$ for $T$, we can compute expectation values for the predictions using $P(\lambda|TD)$. We then find that
 \be
\label{eq:ETtillampt}
{\rm E}[\Ttil(\lambda)] = \Ttil+\overline{\lambda}(T,D)\betatil = \Ttil + \betatil\beta^T(C+S)^{-1}(D-T_0),
\ee
so the predictions are shifted by 
\be 
\label{eq:delTtilpt}
\delta \Ttil = -\Shat(C+S)^{-1}(T_0-D),
\ee
where
\be
\Shat = \betatil\beta^T,\label{eq:Shatdef}
\ee 
is the cross-covariance matrix between observables $T$ for which we have data $D$, and the predictions $\Ttil$. Likewise
\be
\label{eq:varTtillampt}
\Cov[\Ttil(\lambda)] \equiv {\rm E}[(\Ttil(\lambda) -{\rm E}[\Ttil(\lambda)])(\Ttil(\lambda) -E[\Ttil(\lambda)])^T] =\Var[\lambda]\betatil\betatil^T = Z \Stil.
\ee 
so the covariance matrix of the predictions is reduced by the same factor $Z$ as that for the autopredictions.
Thus the data $D$ can lead to more precise (and if the theory is correct, also more accurate) predictions for observables that are not yet measured, through the correlation of theoretical uncertainties. The reduction in the size of the covariance matrix is through the same factor $Z$ as for the autopredictions, while the size of the shift is proportional to the cross-covariance between the theoretical uncertainties.

Note that if, having made the predictions $\Ttil+\delta\Ttil$, an experimentalist made measurements to produce independent data $\widetilde{D}$, with experimental covariance matrix $\widetilde{C}$, and we wanted to combine the datasets $\{D,\widetilde{D}\}$ into a single dataset, then the combined $(N+\widetilde{N})\times(N+\widetilde{N})$ theoretical covariance matrix for $\{T,\widetilde{T}\}$ to be used to compare to the data would be
\be
\left(\begin{array}{cc}
S&\Shat^T\\
\Shat &\Stil\end{array}\right)
 =
\left(\begin{array}{cc}
\beta\beta^T&\beta\betatil^T\\
\betatil\beta^T&\betatil\betatil^T\end{array}\right).
\label{eq:covmatglobal}
\ee
While we might hope that the shifted predictions $\Ttil+\delta\Ttil$ give a better $\chi^2$ to the new data $\widetilde{D}$, this is no longer guaranteed, since the shifts are driven by the old data $D$, and it is possible that $\widetilde{D}$ are inconsistent with them.

\subsection{Autopredictions in Perfect Fits}
\label{subsec:phenomenology}

In the previous section we considered the situation of a theory $T$ which was not fitted to the data $D$. Now consider what is in some sense the opposite situation: a `perfect' fit, where the theoretical predictions $T$ have sufficient flexibility to fit the data $D$ exactly. For a perfect fit, $P(T|D)$  is always maximized when $T=D$, and thus (Eq.~(\ref{eq:chisq})) $\chi^2=0$. We compute expectation values of functions of $T$ using the probability distribution $P(T|D)$ in Eq.~(\ref{eq:PTD}), thus 
\be
\label{eq:ETpheno}
{\rm E}[T] = {\cal N}_T \int dT\; T\; P(T|D) = D,
\ee
the normalization  ${\cal N}_T$ being chosen such that ${\rm E}[1] =1$, while
\be
\label{eq:varTpheno}
\Cov[T] \equiv {\rm E}[(T -{\rm E}[T])(T -{\rm E}[T])^T] = C+S.
\ee
These are the expectation value and covariance matrix of the variables $T$ fitted to the data $D$. 

To compute the autopredictions, we need to consider  
$T(\lambda)=T +\lambda\beta$ (Eq.~(\ref{eq:Tlambda})), and this is more subtle since, as we saw in the previous section, expectation values of $\lambda$ involve $T$. We thus need to generalise the definition Eq.~(\ref{eq:Elampt}) of expectation values of functions of $\lambda$ using the probability distribution $P(\lambda|TD)$, to also include the subsequent integration over $T$: thus we define
\be
\label{eq:Edef}
{\rm E}[f(T,\lambda)] \equiv  {\cal N}_T \int dT\; \Big({\cal N}_\lambda \int d\lambda\;f(T,\lambda)\; P(\lambda|TD)\Big)P(T|D),
\ee
for any function $f(\lambda,T)$ of $\lambda$ and $T$. There are several things to note about this procedure:
\begin{itemize}
\item We always perform the integration over $\lambda$, weighted with the probability distribution $P(\lambda|TD)$, before we perform the integration over $T$ using $P(T|D)$: this is because while $P(\lambda|TD)$ depends on $T$, $P(T|D)$ does not depend on $\lambda$, because it has been marginalised as in Eq.~(\ref{eq:integration}).
\item The data $D$ are always held fixed throughout: both $P(\lambda|TD)$ and $P(T|D)$ are conditional on $D$.
\item For functions $f(\lambda,T)$ which only depend on $T$, the integration over $\lambda$ is trivial and and we recover for example the results
Eqs.~(\ref{eq:ETpheno},\ref{eq:varTpheno}).
\item For the pure theory discussed in the previous section, the theory $T$ was held fixed, so the $T$ integration was trivial and we recover the results Eqs.~(\ref{eq:Elampt},\ref{eq:varlampt}).
\end{itemize}
Thus in the case of a perfect fit, the expectation value of the nuisance parameter is
\be
\label{eq:Elampheno}
{\rm E}[\lambda] = E[\overline{\lambda}(T,D)]=\beta^T(C+S)^{-1}E[D-T] = 0,
\ee
using Eq.~(\ref{eq:lambdabar}) and then finally Eq.~(\ref{eq:ETpheno}). The calculation of the variance requires some care, to ensure that the average over $\lambda$ is separated out from the average over $T$. This is most easily accomplished by adding and subtracting $\overline{\lambda}(T,D)$, so 
\bea
\Var[\lambda] &=& {\rm E}[(\lambda -{\rm E}[\lambda])^2]= {\rm E}[(\lambda -\overline{\lambda}(T,D)+ \overline{\lambda}(T,D))^2]\nn\\
&=& {\rm E}[(\lambda -\overline{\lambda}(T,D))^2]+ E[(\overline{\lambda}(T,D))^2]\nn\\
&=& Z + \beta^T(C+S)^{-1}E[(T-D)(T-D)^T](C+S)^{-1}\beta\nn\\
&=& Z + \beta^T(C+S)^{-1}\Cov[T] (C+S)^{-1}\beta\label{eq:varlamgen}\\
&=& 1 - \beta^T(C+S)^{-1}\beta + \beta^T(C+S)^{-1}\beta = 1.\label{eq:varlampheno}
\eea
where in the second line we note that the cross term vanishes, while in the last line we used Eq.~(\ref{eq:Zdef}) for $Z$, and Eqs.~(\ref{eq:varTpheno}) to simplify the second term. We thus find that in a perfect fit, the probability distribution of the nuisance parameters after fitting the data is the same as the prior distribution Eq.~(\ref{eq:prior}): we learn nothing from the data about the theoretical uncertainty because all the information in the data is absorbed in the fitted parameters. The calculation of the variance is particularly instructive: the reduction by the factor $Z$ found in the pure theory, Eq.~(\ref{eq:varlampt}) is now precisely cancelled by the fluctuation of $\overline{\lambda}(T,D)$ due to the covariance Eq.~(\ref{eq:varTpheno}) of $T$.

Turning to the autopredictions Eq.~(\ref{eq:Tlambda}), we have
\be
\label{eq:ETlampheno}
{\rm E}[T(\lambda)] = E[T+\lambda\beta]=D,
\ee
using Eqs.~(\ref{eq:ETpheno},\ref{eq:Elampheno}): as expected in a perfect fit, the autopredictions simply return the original data. Their covariance is more interesting, as we now have to take account of the correlation between fitted theory $T$ and the nuisance parameter $\lambda$: proceeding as in Eq.~(\ref{eq:varlampheno})
\bea
\label{eq:covTlampheno}
\Cov[T(\lambda)] &=& {\rm E}[(T(\lambda) -{\rm E}[T(\lambda)])(T(\lambda) -{\rm E}[T(\lambda)])^T]\nn\\
&=& {\rm E}[(T -D +\lambda\beta)(T-D+\lambda\beta)^T]\nn\\
&=& {\rm E}[(T-D)(T-D)^T] +  {\rm E}[\lambda \beta(T-D)^T] +  {\rm E}[(T-D)\lambda \beta^T]+ {\rm E}[\lambda^2]\beta\beta^T.
\eea
Now the first term is just $\Cov[T]$, while the last is just $\Var[\lambda]S$, while the cross-terms can be evaluated using Eq.~(\ref{eq:lambdabar}):
\bea
\label{eq:covTlamphenocross} 
{\rm E}[\lambda \beta(T-D)^T] &=& {\rm E}[ \beta\overline{\lambda}(T,D)(T-D)^T] = -S(C+S)^{-1}{\rm E}[(T-D)(T-D)^T] \nn\\
 &=& -S(C+S)^{-1}\Cov[T].
\eea 
We thus find that 
\bea
\Cov[T(\lambda)] &=&  \Cov[T]-S(C+S)^{-1}\Cov[T]-\Cov[T](C+S)^{-1}S+\Var[\lambda]S \label{eq:covTlamgen}\\
&=& (C+S) - S -S + S = C,\label{eq:covTlamphenoC}
\eea
where in the last line we used Eq.~(\ref{eq:varTpheno}) for $\Cov[T]$, and Eq.~(\ref{eq:varlampheno}) for $\Var[\lambda]$. Thus the covariance of the autopredictions in a perfect fit is simply the covariance of the data. The way in which this arises is that the theory covariance arising from the fit (the first term in Eq.(\ref{eq:covTlamgen}) and the theory covariance arising in the autoprediction (the last term in Eq.(\ref{eq:covTlamgen}) are each cancelled by the cross-covariance between fitting and prediction, just as was argued in Ref.~\cite{Harland-Lang:2018bxd}. When we have a perfect fit, there is really no distinction between the autoprediction and the data, and the theory uncertainty thus becomes irrelevant. So in a sense this model is pure phenomenology: the only nontrivial information is in the data. Indeed in this model as it stands there is no possibility of making genuine predictions, since a prediction of the form Eq.~(\ref{eq:Ttillambda}) is useless unless we can determine $\Ttil$, presumably as functions of $T$, and for this we need a genuine theory.

\section{Correlated Theory Uncertainties in One Parameter Fits}
\label{sec:oneparameter}

In the previous section, we considered two simple but unrealistic models: the first in which the theory $T$ is fixed, with no free parameters to be fitted to the data (pure theory), and the second in which the theory $T$ is so flexible that we could achieve a perfect fit, $T=D$ (pure phenomenology). These exercises were useful, in that they gave us some practice in the use of nuisance parameters to propagate theoretical uncertainties. However we now need to consider the more realistic situation in which the theory has parameters that can be constrained by data, but is still sufficiently restrictive that it can be considered a theory. The fit to the data is then not perfect, but the theory is sufficiently constraining that it can be used to predict new observables, $\widetilde T$, for which we as yet have no data.  We will find that the interesting features of the pure theory and pure phenomenology models (the shifts, the reduction in uncertainties due to Bayesian learning, and the correlations between theory uncertainties in the fitting and theory uncertainties in predictions) are also found in the more realistic theories.

In this section we consider a theory with only one fitted parameter: in Sec.~\ref{subsec:single} we explain for the fitting is performed using replicas, in Sec.~\ref{subsec:autoprediction} we consider autopredictions in such a theory, and in Sec.~\ref{subsec:prediction} we consider general predictions. Generalization to many fitting parameters will be considered in the following section.

\subsection{Fitting a Theory with a Single Parameter}
\label{subsec:single}

We can model this situation by considering theoretical predictions $T(\theta)$ which depend on a single parameter $\theta$, so that $\chi^2(\theta)$ is minimized for some choice of this parameter, $\theta=\theta_0$, with some variance $\Var[\theta]$. Other observables $\widetilde{T}(\theta)$ are then predicted to be $\widetilde{T}(\theta_0)$, with an associated uncertainty proportional to $\Var[\theta]$. We assume as before that uncertainties are Gaussian, which means that we can linearize $T(\theta)$ about $T(\theta_0)\equiv T_0$:
\be
\label{eq:Tlin}
T(\theta) = T_0 + (\theta-\theta_0)\Tdot_0.
\ee
This model has the advantage that while it captures the essence of the fitting problem, it is sufficiently simple that we can solve it exactly. 

In order to determine the uncertainty on $\theta$, we will need to propagate the experimental uncertainties in the data $D$ and the theoretical uncertainties in the the predictions $T(\theta)$ into $\theta$. This can be done most easily by generating $N_{\rm rep}$ pseudodata replicas $D^{(r)}$ distributed according to a Gaussian distribution centred on the actual data $D$, with covariance $C+S$: defining the replica average
\be
\label{eq:repav}
\langle F(D^{(r)})\rangle =  \lim_{N_{\rm rep}\to \infty}\smallfrac{1}{N_{\rm rep}}\sum_{r=1}^{N_{\rm rep}}F(D^{(r)})
\ee
for any function $F$ of the replicas, the replicas are chosen such that
\be 
\label{eq:repavD}
\langle D^{(r)}\rangle \equiv D, \qquad \langle (D^{(r)}-D)(D^{(r)}-D)^T\rangle = C+S.
\ee

A parameter replica $\theta^{(r)}$ is then fitted to each pseudodata replica  $D^{(r)}$ 
by maximizing $P(T(\theta)|D^{(r)})$ as given by Eq.~(\ref{eq:PTD}), and thus by 
minimizing  
\be
\label{eq:chi2rep}
\chi_r^2[\theta] = (T(\theta)-D^{(r)})^T(C+S)^{-1}(T(\theta)-D^{(r)}),
\ee
with respect to $\theta$, replica by replica. Using Eq.~(\ref{eq:Tlin}), minimization of the quadratic gives
\be
\label{eq:arep}
\theta^{(r)} - \theta_0 = \frac{\Tdot_0^T(C+S)^{-1}(D^{(r)}-T_0)}{\Tdot_0^T(C+S)^{-1}\Tdot_0}.
\ee
Using the replica averages Eq.~(\ref{eq:repavD}), and choosing $\theta_0 = \langle \theta^{(r)}\rangle$, we find for consistency
\be
\label{eq:consistency}
\Tdot_0^T(C+S)^{-1}(D-T_0)=0,
\ee
and thus we can rewrite Eq.~(\ref{eq:arep}) as 
\be
\label{eq:arep2}
\theta^{(r)} - \theta_0 = \frac{\Tdot_0^T(C+S)^{-1}(D^{(r)}-D)}{\Tdot_0^T(C+S)^{-1}\Tdot_0}.
\ee
Then since $C$ and $S$ are symmetric matrices,
\bea
\Var[\theta] &=& \langle(\theta^{(r)}-\theta_0)^2\rangle\nn\\
 &=& \frac{\Tdot_0^T(C+S)^{-1}\langle(D^{(r)}-D)(D^{(r)}-D)^T\rangle (C+S)^{-1}\Tdot_0}{(\Tdot_0^T(C+S)^{-1}\Tdot_0)^2}\nn\\
&=& (\Tdot_0^T(C+S)^{-1}\Tdot_0)^{-1}.
\label{eq:vara}
\eea
Note the way the double reciprocation in this expression works: data points with a relatively large dependence on $\theta$ (i.e. large $\Tdot_0$) contribute more than those with small dependence, however directions with large uncertainty (i.e. projections of $C+S$) contribute less than those with small uncertainty.

Now that we understand the uncertainty of the fitted parameter $\theta$, we can use it to predict the uncertainties on $T(\theta)$: 
\be
\label{eq:repET}
E[T] \equiv \langle T(\theta^{(r)})\rangle = T(\theta_0) = T_0,
\ee
so, writing $T^{(r)} = T(\theta^{(r)})$, 
\bea
X\equiv\Cov[T(\theta)] &=& \langle(T^{(r)}-T_0)(T^{(r)}-T_0)^T\rangle\label{eq:Xdef}\\
&=& \Tdot_0\langle(\theta^{(r)}-\theta_0)^2\rangle\Tdot_0^T
= \Tdot_0(\Tdot_0^T(C+S)^{-1}\Tdot_0)^{-1}\Tdot_0^T\label{eq:Xdef2}\\
&=& n(n^T(C+S)^{-1}n)^{-1}n^T,
\label{eq:Xdef3}
\eea
where $n$ is a unit vector in the direction of $\Tdot_0$, $n^Tn=1$. This shows that $X$ depends only on $n$, and 
not on $|\Tdot_0|$. 

The singular matrix $X$ will play an important role in what follows: it is the covariance matrix of $T$ due to the experimental and theoretical uncertainties in the fitting of the parameter $\theta$ --- the `fitting uncertainty'. When the fitted parameter minimizes the $\chi^2$, and is thus given by Eq.~(\ref{eq:arep2}), $X$ satisfies the projective relation
 \be
\label{eq:XsqeqX}
X = X(C+S)^{-1}X.
\ee
Using Eq.~(\ref{eq:arep2}) in Eq.~(\ref{eq:Tlin}), we see that $X(C+S)^{-1}$ projects the data replicas onto the theory replicas:
\be
T^{(r)}-T_0 = X(C+S)^{-1}(D^{(r)}-D).
\label{eq:projection}
\ee
Because this relation is projective, some information is lost whenever we perform the fit: Eq.~(\ref{eq:projection}) cannot be inverted to obtain data replicas from theory replicas. This is an inevitable consequence of describing $N$ data (assuming $N>1$) with only a single parameter $\theta$.

To make $X$ more explicit, consider the simple model Eq.~(\ref{eq:modelCS}) for $C$ and $S$. Then using Eq.~(\ref{eq:modelCplusSinv}), 
\be
\label{eq:denom}
n^T(C+S)^{-1}n= \frac{\sigma^2+s^2\sin^2\phi}{\sigma^2(\sigma^2+s^2)},
\ee
where $\cos\phi = n^Te$, and thus if we project $X$ onto $n$ 
(projections orthogonal to $n$ give zero)
\be
\label{eq:nXnmod}
n^TXn = \frac{\sigma^2(\sigma^2+s^2)}{(\sigma^2+s^2\sin^2\phi)}.
\ee
The contribution of the theory uncertainty $s$ thus depends on how well aligned $e$ is to the direction $n$ of the parameter dependence: if $\phi=0$ we have complete alignment, and the variance of $T$ in this direction is $\sigma^2+s^2$ as expected, while if $\phi=\frac{\pi}{2}$ we have orthogonality, and the variance of $T$ is $\sigma^2$ --- the theory uncertainty is then irrelevant to the fitting. 

\subsection{Autopredictions in Single Parameter Fits}
\label{subsec:autoprediction}

We can now consider the evaluation of the mean and covariance of the `autopredictions'  
\be
\label{eq:autopred}
T(\theta,\lambda)=T(\theta)+\lambda\beta 
\ee
in this one parameter model. Just as in Sec.~\ref{subsec:phenomenology}, we do this by first computing expectation values over $\lambda$, using $P(\lambda|TD)$, which depend on $T$, and then evaluate the expectation values over $T$, according to the probability distribution $P(T|D)$, now performed by averaging over theory replicas $T^{(r)} = T(\theta^{(r)})$. It is important to note that both these averages are performed holding the data $D$ fixed, as both probabilities are conditional on the data: the data replicas $D^{(r)}$ employed in Sec.~\ref{subsec:single} are only a device to generate the theory replicas $T^{(r)}$, and are not to be averaged over when determining expectation values. Accordingly, Eq.~(\ref{eq:Edef}) now becomes
\be
\label{eq:Edef}
{\rm E}[f(T,\lambda)]  =\Big\langle \Big({\cal N}_\lambda \int d\lambda\;f(T^{(r)},\lambda)\; P(\lambda|T^{(r)}D)\Big)\Big\rangle,
\ee  
where the angled brackets denote the replica average Eq.~(\ref{eq:repav}).

Following the same steps as in the perfect fit in Sec.~\ref{subsec:phenomenology}, but now using the theory replicas $T^{(r)} = T(\theta^{(r)})$ determined in the one parameter fit Sec.~\ref{subsec:single}, we find
\be
\label{eq:explam}
E[\lambda] \equiv \langle \overline\lambda(T(\theta^{(r)}),D)\rangle = \beta^T(C+S)^{-1}(D-T_0)\equiv\overline\lambda_0,
\ee
Unlike in the perfect fit Eq.~(\ref{eq:Elampheno}), but just as in the pure theory Eq.~(\ref{eq:Elampt}) the nuisance parameters can thus now have nonzero expectation values, since the one parameter fit no longer fits the data exactly. These in turn give nontrivial shifts in the theoretical predictions: 
\be
\label{eq:ET}
E[T(\theta,\lambda)] = \langle T^{(r)}+\overline\lambda(T^{(r)},D)\beta\rangle   
 = T_0+\overline\lambda_0\beta = T_0+\beta\beta^T(C+S)^{-1}(D-T_0).
\ee
So again the data give us information, inducing shifts in the autopredictions:
\be
\label{eq:shift}
\delta T  = -S(C+S)^{-1}(T_0-D).
\ee
Note however that since (from Eq.~(\ref{eq:consistency})) $n^T(C+S)^{-1}(T_0-D)=0$, these shifts will only be nonzero when $n$ and $e$ (the data and the theory) point in different directions: when they are parallel ($\phi=0$), the theoretical uncertainty is simply absorbed by the fit, just as it was in the perfect fit in Sec.~\ref{subsec:phenomenology}. We can use the same argument as in Sec.~\ref{subsec:puretheory}, Eq.~(\ref{eq:chisqauto}), to show that the shifts will always improve the fit to the experimental data.

For the uncertainties, consider first the variance of $\lambda$: following the same argument that led to Eq.~(\ref{eq:varlampheno}) in Sec.~{subsec:phenomenology}, we now
find
\bea
\Var[\lambda] &=& {\rm E}[(\lambda -{\rm E}[\lambda])^2]= {\rm E}[(\lambda -\overline{\lambda}(T,D)+ \overline{\lambda}(T,D)-\overline\lambda_0)^2]\nn\\
&=& {\rm E}[(\lambda -\overline{\lambda}(T,D))^2]+ \langle(\overline{\lambda}(T^{(r)},D)-\overline\lambda_0)^2\rangle\nn\\
&=& Z + \beta^T(C+S)^{-1}\langle(T^{(r)}-T_0)(T^{(r)}-T_0)^T\rangle(C+S)^{-1}\beta\nn\\
&=& 1 - \beta^T(C+S)^{-1}\beta + \beta^T(C+S)^{-1}X(C+S)^{-1}\beta\equiv \Zbar .\label{eq:Zbardef}
\eea
where in the second line we turned the expectation value over $T$ in to a replica average, and in the last line we used Eq.~(\ref{eq:Zdef}) for $Z$, and Eq.~(\ref{eq:Xdef}) for $\Cov[T]$. We thus find that in the more restrictive environment of the one parameter fit, the last two terms no longer cancel: the information in the data can no longer be entirely absorbed in the single fitted parameter, and so it can still inform the nuisance parameter.  

It is easy to see that $\Zbar\geq Z$ because $(C+S)^{-1}X(C+S)^{-1}$ is positive semi-definite, while $\Zbar\leq 1$ since $X(C+S)^{-1}$ 
is projective, Eq.~(\ref{eq:XsqeqX}, so its eigenvalues are either zero or one. So in place of Eq.~(\ref{zbounds}) we now have
\be
\label{Zbarbounds}
0<Z\leq\Zbar\leq 1.
\ee
The information on theoretical uncertainties extracted from the data is thus less in the one parameter fit than it was in the pure theory of Sec.~\ref{subsec:puretheory}, due to the extra uncertainty arising in the fit itself, but unlike in the perfect fit Sec.~\ref{subsec:phenomenology}, the data will still constrain the theoretical uncertainties provided the parameter and theoretical uncertainty act in different directions.

In the model Eq.~(\ref{eq:modelCS}) for $C$ and $S$, and using Eq.~(\ref{eq:nXnmod}) for $X$,  Eq.~(\ref{eq:Zbardef}) becomes 
\be
\label{eq:Zbarmod}
\Zbar = \frac{\sigma^2}{\sigma^2+s^2\sin^2\phi}.
\ee
Comparing with the corresponding expression for $Z$, Eq.~(\ref{eq:modelZ}), we see that indeed $\Zbar=1$ when $\phi=0$, thus when $n=e$ and the parameter variation and theoretical uncertainty are aligned, while $\Zbar=Z$ only if $\phi=\pi/2$, so when $n$ and $e$ are orthogonal, the data have the greatest influence on the uncertainty.

For the covariance of the autopredictions Eq.~(\ref{eq:autopred}) we also have to take account of the correlation between the fitted theory and the nuisance parameter: proceeding as in Eq.~(\ref{eq:covTlampheno})
\bea
\Cov[T(\theta,\lambda)] &=& {\rm E}[(T(\theta,\lambda) -{\rm E}[T(\theta,\lambda)])(T(\theta,\lambda) -{\rm E}[T(\theta,\lambda)])^T]\nn\\
&=& {\rm E}[(T -T_0 +(\lambda-\overline\lambda_0)\beta)(T-T_0+(\lambda-\overline\lambda_0)\beta)^T]\nn\\
&=& \langle(T^{(r)}-T_0)(T^{(r)}-T_0)^T\rangle +  {\rm E}[(\lambda-\overline\lambda_0) \beta(T-T_0)^T] \nn\\
&&\qquad +  {\rm E}[(T-T_0)(\lambda-\overline\lambda_0) \beta^T]+ {\rm E}[(\lambda -\overline\lambda_0)^2]\beta\beta^T.\label{eq:covTlamonex}
\eea
Then again the first term is $\Cov[T]=X$, Eq.~(\ref{eq:Xdef}), while the last is $\Var[\lambda]S$, Eq.~(\ref{eq:Zbardef}), while the cross-terms can be evaluated using Eq.~(\ref{eq:lambdabar}):
\bea
\label{eq:covTlamphenocross} 
{\rm E}[(\lambda-\overline\lambda_0) \beta(T-T_0)^T] &=& \langle \beta(\overline{\lambda}(T^{(r)},D)-\overline{\lambda}(T_0,D)(T^{(r)}-T_0)^T\rangle \nn\\
&=& -S(C+S)^{-1}\langle(T^{(r)}-T_0)(T^{(r)}-T_0)^T\rangle \nn\\
 &=& -S(C+S)^{-1}\Cov[T].
\eea 
We thus find that the result is simply 
\be
\Cov[T(\lambda)] =  X-S(C+S)^{-1}X-X(C+S)^{-1}S+\Zbar S. \label{eq:covTlamone}
\ee
The meaning of the four terms is easy to understand: the first is the `fitting uncertainty' (which includes contributions from both experimental and theoretical uncertainties), the last the `theory uncertainty', reduced through exposure to the data, and the middle two terms are due to the correlations between the two sources of theoretical uncertainty. 
We can simplify it by using Eq.~(\ref{eq:Zbardef}) to show that
\be
\label{eq:Zbaralgebra}
\Zbar S = S(C+S)^{-1}X(C+S)^{-1}S + ZS,
\ee
and then some straightforward algebra to write
\be
\label{eq:Xalgebra}
X-S(C+S)^{-1}X-X(C+S)^{-1}S+S(C+S)^{-1}X(C+S)^{-1}S = C(C+S)^{-1}X(C+S)^{-1}C.
\ee
We thus find finally
\be
\Cov[T(\lambda)] =   C(C+S)^{-1}X(C+S)^{-1}C + ZS.\label{eq:covTlamfin}
\ee
Note that if we write $X=C+S$ (as in the perfect fit model in Sec.~\ref{subsec:phenomenology}), this result reduces to $C$, as it should. The cancellations noted in Eq.~(\ref{eq:covTlamphenoC}) between the cross-terms and the covariances of of $T$ and $\lambda$ are no longer exact in Eq.~(\ref{eq:covTlamfin}) , because $\Cov[T]$ is no longer $C+S$, but rather the smaller matrix $X$ (which is in a sense $C+S$ restricted to the space of variation of the parameter $\theta$ as in Eq.~~(\ref{eq:Xdef})). Thus the result is no longer the experimental covariance matrix $C$, but rather the sum in quadrature of the `fitting uncertainty', $X$, and the `theory uncertainty' $S$, each reduced to some extent by the correlation of the theoretical uncertainties in fit and prediction. 

In the model Eq.~(\ref{eq:modelCS}) for $C$ and $S$, and using Eq.~(\ref{eq:nXnmod}) for $X$, we now find
\be
\Cov[T(\lambda)] =   \frac{\sigma^2(\sigma^2+s^2)}{(\sigma^2+s^2\sin^2\phi)}
\Big(nn^T - \frac{s^2}{\sigma^2+s^2}\cos\phi(en^T+ne^T) + \frac{s^2}{\sigma^2+s^2}ee^T\Big) .\label{eq:covTlamonemod}
\ee
The first term is just $X$, the off-diagonal term in the middle is the correlation term, and the last is $\Zbar S$. If $\phi=0$, so $n=e$, the three terms combine to give simply $\sigma^2nn^T$: in the direction of the fitted parameter, the uncertainty in the autoprediction is the experimental uncertainty, just as in the perfect fit described in Sec.~\ref{subsec:phenomenology}. On the other hand, if $\phi=\pi/2$, so $n$ and $e$ are orthogonal, the correlation term disappears, and the result reduces to $X+ZS$: we add the uncertainties in quadrature, since they are in orthogonal directions, and the theoretical uncertainty is reduced just as in the pure theory described in Sec.~\ref{subsec:puretheory}. The one parameter fit thus interpolates smoothly between these two extremes. Note that we can write Eq.~(\ref{eq:covTlamonemod}) as 
\be
\Cov[T(\lambda)] =   \frac{\sigma^2(\sigma^2+s^2)}{(\sigma^2+s^2\sin^2\phi)}
\Big(n - \frac{s^2\cos\phi}{\sigma^2+s^2} e\Big)\Big(n^T - \frac{s^2\cos\phi}{\sigma^2+s^2}e^T\Big) + \frac{s^2\sigma^2}{\sigma^2+s^2}ee^T .\label{eq:covTlamonemodx}
\ee
Here the last term is just $ZS$, while the first is $X$, but with the vector $n$ given an additional component in the direction $e$ due to the theory correlation: besides changing its direction, this reduces the size of the fitting uncertainty by a factor $\sqrt{\sin^2\phi + Z^2\cos^2\phi}$. However it is easy to see that the size of the correlated fitting uncertainty is still larger than it would be if the theory uncertainty had not been included in the fit.

\subsection{Correlated Predictions in One Parameter Fits}
\label{subsec:prediction}

We now consider predictions $\Ttil_I(\theta)$, $I=1,\ldots,\widetilde{N}$, which depend on the same parameter $\theta$ as the fitted predictions $T_i(\theta)$, $i=1,\ldots, N$. There are two distinct sources of uncertainty in $\Ttil_I(\theta)$: uncertainties in the determination of $\theta$ due to the experimental uncertainties in the data $D_i$ and  theoretical uncertainties in the theory $T_i(\theta)$ used in its determination; and theoretical uncertainties in the predictions $\Ttil_I(\theta)$ themselves. 

The first uncertainty is expressed through Eq.~(\ref{eq:arep2}), which gives the variance Eq.~(\ref{eq:vara}). In analogy with Eq.~(\ref{eq:Tlin}) the linearised dependence of the predictions $\Ttil(\theta)$ may be written
\be
\label{eq:Tlin2}
\Ttil(\theta) = \Ttil_0 + (\theta-\theta_0)\Ttildot_0,
\ee
with $\Ttil_0=\Ttil(\theta_0)$. Then the covariance of $\Ttil_I(\theta)$ due to the uncertainty in the parameter $\theta$ is derived just as in Eqs.~(\ref{eq:Xdef}-\ref{eq:Xdef3}) for the autopredictions: writing $\Ttil^{(r)}\equiv\Ttil(\theta^{(r)})$
\bea
\Xtil\equiv\Cov[\Ttil(\theta)] &=& \langle(\Ttil^{(r)}-\Ttil_0)(\Ttil^{(r)}-\Ttil_0)^T\rangle\label{eq:Xtildef}\\
&=& \Ttildot_0\langle(\theta^{(r)}-\theta_0)^2\rangle\Ttildot_0^T
= \Ttildot_0(\Tdot_0^T(C+S)^{-1}\Tdot_0)^{-1}\Ttildot_0^T\label{eq:Xtildef2}.
\eea

The second uncertainty --- the theoretical uncertainty in the predictions $\Ttil_I(\theta)$ --- may again be either correlated or uncorrelated with the theoretical uncertainty in $T_i(\theta)$. Consider first the simpler situation when it is uncorrelated: this might be the case if, for example, the observable $\Ttil(\theta)$ was a different type of observable to the $T(\theta)$ used to determine $\theta$. Then 
introducing a nuisance parameter $\lambdatil$, Gaussian distributed about zero with unit variance, and uncorrelated with $\lambda$, we can write (as in the pure theory model Eq.~(\ref{eq:Ttillambda}))
\be
\Ttil(\theta,\lambdatil) =  \Ttil(\theta) + \lambdatil\betatil,\label{eq:uncor}
\ee
where the vector $\betatil_I$ gives the size of the theoretical uncertainties
in $\Ttil_I(\theta)$.
Since $\theta$ and $\lambdatil$ are uncorrelated, we then have
\bea
E[\Ttil(\theta,\lambdatil)] &=& \Ttil(\theta_0),\label{eq:uncorpred}\\
\Cov[\Ttil(\theta,\lambdatil)]&=& \Cov[\Ttil(\theta)]+\Var[\lambdatil]\betatil\betatil^T 
= \Xtil + \Stil,\label{eq:uncorcov}
\eea
where $\Stil = \betatil\betatil^T$
is the theory covariance matrix for the prediction $\Ttil(\theta)$ (compare Eq.~(\ref{eq:Stildef}) in the pure theory). Thus when the theoretical uncertainty is uncorrelated we simply add it in quadrature to the uncertainty due to that in the parameter $\theta$ derived from the fit.

Now consider the more interesting case in which the theoretical uncertainty in 
$\Ttil_I(\theta)$ is fully correlated to that in the $T_i(\theta)$ used in the fit to determine $\theta$: then $\lambdatil=\lambda$, which has already been determined in the fit to have nonzero expectation value and variance Eqs.~(\ref{eq:explam},\ref{eq:Zbardef}).
Then writing $\Ttil(\theta,\lambda) = \Ttil(\theta)+\lambda\betatil$,
\be
\label{eq:corpredmean}
E[\Ttil(\theta,\lambda)] = \Ttil_0 + \overline\lambda(T_0,D)\betatil,
\ee
so the correlation induces a similar shift in the predictions to that in the autopredictions Eq.~(\ref{eq:shift}): using Eq.~(\ref{eq:lambdabar}) 
\be
\label{eq:shiftpred}
\delta \Ttil(\theta_0) = \betatil\beta^T(C+S)^{-1}(D-T_0) = -\Shat (C+S)^{-1}(T_0-D).
\ee
where $\Shat = \betatil\beta^T$, Eq.~(\ref{eq:Shatdef}) is the matrix of cross-correlations between observables $T(\theta)$ used in the fit and the predictions $\Ttil(\theta)$ .

Likewise using the same arguments as were used to derive $\Cov[T(\theta,\lambda)]$, Eq.~(\ref{eq:covTlamone})
\bea
\Cov[\Ttil(\theta,\lambda)] &=& {\rm E}[(\Ttil(\theta,\lambda) -{\rm E}[\Ttil(\theta,\lambda)])(\Ttil(\theta,\lambda) -{\rm E}[\Ttil(\theta,\lambda)])^T]\nn\\
&=& \langle\Ttil^{(r)}-\Ttil_0)(\Ttil^{(r)}-\Ttil_0)^T\rangle +  {\rm E}[(\lambda-\overline\lambda_0) \betatil(\Ttil-\Ttil_0)^T] \nn\\
&&\qquad +  {\rm E}[(\Ttil-\Ttil_0)(\lambda-\overline\lambda_0) \betatil^T]+ {\rm E}[(\lambda -\overline\lambda_0)^2]\betatil\betatil^T.\label{eq:covTtillamonex}
\eea
Then again the first term is $\Cov[\Ttil]=\Xtil$, Eq.~(\ref{eq:Xtildef}), while the last is $\Var[\lambda]\Stil$, Eqs.~(\ref{eq:Zbardef},\ref{eq:Stildef}), while the cross-terms can be evaluated using Eq.~(\ref{eq:lambdabar}):
\bea
\label{eq:covTtillamphenocross} 
{\rm E}[(\lambda-\overline\lambda_0) \betatil(\Ttil-\Ttil_0)^T] &=& \langle \betatil(\overline{\lambda}(T^{(r)},D)
-\overline{\lambda}(T_0,D))(\Ttil^{(r)}-\Ttil_0)^T\rangle \nn\\
&=& -\Shat(C+S)^{-1}\langle(T^{(r)}-T_0)(\Ttil^{(r)}-\Ttil_0)^T\rangle\nn\\ &=& -\Shat(C+S)^{-1}\Xhat^T,
\eea
where $\Shat$ is given by Eq.~(\ref{eq:Shatdef}), and in analogy to Eq.~(\ref{eq:Xdef}) and Eq.~(\ref{eq:Xtildef}) we define the cross-covariance between $\Ttil$ and $T$
\bea
\Xhat &\equiv& \langle(\Ttil^{(r)}-\Ttil_0)(T^{(r)}-T_0)^T\rangle\label{eq:Xhatdef}\\
&=& \Ttildot_0\langle(\theta^{(r)}-\theta_0)^2\rangle\Tdot_0^T=\Ttildot_0(\Tdot_0^T(C+S)^{-1}\Tdot_0)^{-1}\Tdot_0^T\label{eq:Xhatdef2}.
\eea 
We thus find that  
\be
\Cov[\Ttil(\theta,\lambda)] =  \Xtil-\Shat(C+S)^{-1}\Xhat^T-\Xhat(C+S)^{-1}\Shat^T+\Zbar \Stil. \label{eq:covTtillamone}
\ee
Using Eq.~(\ref{eq:Zbardef}), we can write the last term as
\bea
\Zbar \Stil &=& Z\Stil + \Shat(C+S)^{-1}X(C+S)^{-1}\Shat^T,\label{eq:ZbarStil}\\
Z\Stil &=&\Stil  - \Shat(C+S)^{-1}\Shat^T. 
\label{eq:ZStil}
\eea
Note that the coefficients $Z$ and $\Zbar$ are the same as for the autopredictions, and thus satisfy the bounds Eq.~(\ref{Zbarbounds}), we must have (for positive definite $C$ and positive semi-definite $S$, i.e. $C>0$, $S\geq 0$) 
\be
\label{eq:possemdeftil}
0\leq \Shat(C+S)^{-1}X(C+S)^{-1}\Shat^T\leq \Shat(C+S)^{-1}\Shat^T \leq \Stil,
\ee
so in particular the subtraction (the last term in Eq.~(\ref{eq:ZStil}) can never be so large that it makes the entire covariance matrix negative.  

In summary, comparing Eqs.~(\ref{eq:shiftpred},\ref{eq:covTtillamone}) with 
Eqs.~(\ref{eq:uncorpred},\ref{eq:uncorcov}), we see that including the correlations between the theoretical uncertainties in the fit and the prediction results in three effects: a shift in the central value of the prediction, a reduction in the theoretical uncertainty, and a reduction in the fitting uncertainty due to the correlations. Performing the fit gives us information (from the data) about the theory, which results in more precise, and hopefully more accurate, predictions.

\section{Correlated MHOU in PDF fits}
\label{sec:corrlnpdffits}

We now repeat the above analysis, but instead of the toy model we consider the more realistic situation in which the theoretical expressions $T_i[f]$ depend on PDFs $f$, determined in a global fit to $N$ data $D_i$, with experimental covariance matrix $C_{ij}$, and then used to make $\widetilde{N}$ predictions $\Ttil_I[f]$. There are then many sources of theoretical uncertainty in the relation between the theoretical calculations and the PDFs: here we consider the most generic, the missing higher order uncertainty (MHOU), computed using scale variations according to one of the prescriptions set out in Ref.~\cite{AbdulKhalek:2019bux,AbdulKhalek:2019ihb}. The theory covariance matrices $S_{ij}$ and $\Stil_{IJ}$ associated with the MHOU will then have many non-zero eigenvalues, and thus there will be $n$ nuisance parameters $\lambda_\alpha$, $\alpha=1,\ldots, n$ to take into account. There is in principle no limit on $n$, though in practice $n\ll N$. As in the toy model, the fit to the PDFs will determine the mean and covariance of the nuisance parameters, which will then translate into systematic shifts and changes in the uncertainties of the theoretical predictions. 

\subsection{Expectation and Covariance of Multiple Nuisance Parameters}
\label{subsec:multiplenuisance}

The nuisance parameters $\lambda_\alpha$ correspond to shifts in the theoretical predictions: $T_i[f]\to T_i[f] + \lambda_\alpha\beta_{i,\alpha}[f]$, where we adopt the summation convention for the index $\alpha$. The shift vectors $\beta_{i,\alpha}$ are not necessarily orthogonal to each other. We again assume Gaussian uncertainties, so that in place of Eq.~(\ref{eq:PTDl}) we now have
\be
\label{eq:PTDlf}
P(T|D\lambda)\propto \exp\big(-\half(T[f]+\lambda_\alpha\beta_\alpha-D)^TC^{-1}(T[f]+\lambda_\alpha\beta_\alpha-D)\big),
\ee
and assume that each nuisance parameter has a prior which is Gaussian distributed with unit variance, centred on zero, the distributions being independent both of each other and of the data, so that 
\be
\label{eq:priorf}
P(\lambda|D)=P(\lambda) \propto \exp\big(-\half\lambda_\alpha\lambda_\alpha\big).
\ee
We now marginalize over $\lambda_\alpha$, as in Eq.~(\ref{eq:marginalise1})
\be
\label{eq:marginalise2}
P(T|D) \propto\int d^n\lambda\, \exp\left(-\half[(T[f]+\lambda_\alpha\beta_\alpha-D)^TC^{-1}(T[f]+\lambda_\beta\beta_\beta-D)+\delta_{\alpha\beta}\lambda_\alpha\lambda_\beta]\right)\, , 
\ee
by first completing the square: the details are messy, but the result is very similar to Eq.~(\ref{eq:integration}), namely
\be
\label{eq:integrationf}
P(T|D)\propto\int d^n\lambda\, \exp\left(-\half(\lambda_\alpha-\overline{\lambda}_\alpha) Z_{\alpha\beta}^{-1}(\lambda_\beta-\overline{\lambda}_\beta) - \half\chi^2\right) \propto \exp(-\half\chi^2),
\ee
 where now 
\be
\label{eq:Zdefmat}
Z_{\alpha\beta} = (\delta_{\alpha\beta}+\beta_\alpha^TC^{-1}\beta_\beta)^{-1},
\ee
the inverse on the right hand side being the matrix inverse with respect to the indices $\alpha$ and $\beta$,
\be
\label{eq:lambdabarf}
\overline{\lambda}_\alpha(T,D) = Z_{\alpha\beta}\beta_\beta^TC^{-1}(D-T),
\ee
and $\chi^2$ is once again given by Eq.~(\ref{eq:chisq}), but now with in place of Eq.~(\ref{eq:Sdef})
\be
\label{eq:Sdeff}
S = \beta_\alpha\beta^T_\alpha,
\ee
as expected. The Gaussian integration in Eq.~(\ref{eq:integrationf}) is now trivial, taking us back again to Eq.~(\ref{eq:PTD}) up to a factor $(2\pi)^{n/2}({\rm det}Z)^{1/2}$, which we can ignore as it does not depend on $T$ or $D$, while Bayes' Theorem Eq.~(\ref{eq:bayes}) gives us the posterior distribution of the nuisance parameters:  
\be
\label{eq:posteriorf}
P(\lambda|TD)\propto \exp\big(-\half(\lambda_\alpha-\overline{\lambda}_\alpha) Z_{\alpha\beta}^{-1}(\lambda_\beta-\overline{\lambda}_\beta)\big),
\ee
whence we see that 
\be
\label{eq:meanvarlamf}
E[\lambda_\alpha] =\lambdabar_\alpha,\qquad E[(\lambda_\alpha-\lambdabar_\alpha)(\lambda_\beta-\lambdabar_\beta)] = Z_{\alpha\beta}.
\ee
 It is easy to see from the definition Eq.~(\ref{eq:Zdefmat}) that if $e_\alpha$ is a unit eigenvector of $Z_{\alpha\beta}$, and $\beta=e_\alpha\beta_\alpha$, then the corresponding eigenvalue of $Z_{\alpha\beta}$ is $z = (1+\beta^TC^{-1}\beta)^{-1}$, so $0<z<1$, and (in analogy to the bounds Eq.~(\ref{zbounds}) $Z_{\alpha\beta}$ is positive definite (thus invertible) and $\delta_{\alpha\beta}-Z_{\alpha\beta}$ is also positive definite (because the eigenvalues $z$ are all less than one). We can summarise this by writing, in place of Eq.~(\ref{zbounds})
\be
0 < Z_{\alpha\beta} < \delta_{\alpha\beta}.
\label{eq:Zabbounds}
\ee

We can express $Z_{\alpha\beta}$ in terms of the inverse of $C+S$:
\be
\label{eq:Zdefab}
Z_{\alpha\beta} = \delta_{\alpha\beta}-\beta_\alpha^T(C+S)^{-1}\beta_\beta,
\ee
since 
\bea
\label{eq:invalgebra}
&&(\delta_{\alpha\gamma}+\beta_\alpha^TC^{-1}\beta_\gamma)(\delta_{\gamma\beta}-\beta_\gamma^T(C+S)^{-1}\beta_\beta)\nn\\
&&\qquad\qquad\qquad=\delta_{\alpha\beta} + \beta_\alpha^T(C^{-1}-(C+S)^{-1}-C^{-1}S(C+S)^{-1})\beta_\beta = \delta_{\alpha\beta}.
\eea
Combining Eq.~(\ref{eq:Zdefab}) with Eq.~(\ref{eq:lambdabarf}), we then have
\be
\label{eq:lambdabarfx}
\overline{\lambda}_\alpha = \beta_\alpha^T(C+S)^{-1}(D-T),
\ee
since $(1-(C+S)^{-1}S)C^{-1} = (C+S)^{-1}$.

\subsection{Fitting the PDFs}
\label{subsec:pdfexactfit}

We now proceed to apply the above results in the context of a PDF fit incorporating MHOU~\cite{AbdulKhalek:2019bux,AbdulKhalek:2019ihb}. We consider first a fixed parametrization: the PDFs $f(\theta)$ will then depend on $m$ parameters $\theta_p$, $p = 1,\ldots,m$, with $m<N$ so the data $D$ are sufficient to determine all the parameters through minimization of the $\chi^2$ 
Eq.~(\ref{eq:chisq}).
 
We can then follow the same procedure as in Sec.\ref{sec:oneparameter}, the only difference being that now we fit the $m$ parameters $\theta_p$ rather than just the single parameter $\theta$. Writing the theoretical predictions $T[f(\theta)]\equiv T(\theta)$, the linearization relation Eq.~(\ref{eq:Tlin2}) becomes
\be
\label{eq:Tlinf}
T(\theta) = T_0 + (\theta_p-\theta_p^0)T_p,
\ee
where $f(\theta^0)\equiv f_0$ is the PDF that minimizes the $\chi^2$, $T_0\equiv T(\theta^0)$, $T_p\equiv \partial T(\theta^0)/\partial\theta_p^0$, and we use the summation convention for indices $p$. Using the data replicas Eq.~(\ref{eq:repav},\ref{eq:repavD}), minimizing Eq.~(\ref{eq:chi2rep}) with respect to $\theta_p$ rather than $\theta$, we find in place of  Eq.~(\ref{eq:arep2}) that the 
fluctuations of the PDF replica parameters are given by
\be
\label{eq:arep2f}
\theta^{(r)}_p - \theta^0_p = (T_p^T(C+S)^{-1}T_q)^{-1}T_q^T(C+S)^{-1}(D^{(r)}-D),
\ee
where the matrix inverse in the first factor on the right hand side is with respect to the $p,q$ indices. It follows that instead of Eq.~(\ref{eq:vara}) we now have the covariance matrix
\bea
\Cov_{pq}[\theta] &=& \langle(\theta^{(r)}_p-\theta_p^0)(\theta^{(r)}_q-\theta_q^0)\rangle\nn\\
&=& (T_p^T(C+S)^{-1}T_q)^{-1}, 
\label{eq:varaf}
\eea
while the expression Eq.~(\ref{eq:Xdef3}) for the covariance of the predictions $T[f]$ becomes, on writing $T_p = |T_p|n_p$, where $n_p$ are unit vectors (which are however not necessarily orthogonal) 
\be
X = n_p(n_p^T(C+S)^{-1}n_q)^{-1}n_q^T.
\label{eq:Xdeffpq}
\ee
so the projective relation Eq.~(\ref{eq:XsqeqX}) still holds, and $X(C+S)^{-1}$ projects data replicas onto theory replicas as in Eq.~(\ref{eq:projection}).

It is now easy to see that the results for the autopredictions in Sec.\ref{subsec:autoprediction} continue to hold, and that in particular that since the central values of the nuisance parameters 
$\overline{\lambda}_\alpha$ are given by
\be
\label{eq:Elambdaf}
{\rm E}[\lambda_\alpha] = -\beta_\alpha^T(C+S)^{-1}(\langle T^{(r)}\rangle-D),
\ee
the shifts Eq.~(\ref{eq:shift}) are now
\be
\label{eq:shiftmult}
\delta T[f] = \beta_\alpha\beta_\alpha^T(C+S)^{-1}(D-T[f_0]) = -S(C+S)^{-1}(T[f_0]-D).
\ee
These shifts will improve the $\chi^2$ to the experimental data, in just the same way as in Sec.~\ref{subsec:puretheory}.

Likewise for the uncertainties: Eq.~(\ref{eq:Zbardef}) for the variance of the nuisance parameter becomes an equation for the covariance matrix of the nuisance parameters in the context of the PDF fit,
\bea
\Cov_{\alpha\beta}[\lambda] &\equiv& {\rm E}[(\lambda_\alpha - {\rm E}[\lambda_\alpha])(\lambda_\beta - {\rm E}[\lambda_\beta])]\nn\\
&=&\delta_{\alpha\beta} -\beta_\alpha^T(C+S)^{-1}\beta_\beta-\beta_\alpha^T(C+S)^{-1}X(C+S)^{-1}\beta_\beta\equiv \Zbar_{\alpha\beta},
\label{eq:Zbardefab}\eea
using the projective relation Eq.~(\ref{eq:XsqeqX}). Again, both $\Zbar_{\alpha\beta}-Z_{\alpha\beta}$ and $\delta_{\alpha\beta}-\Zbar_{\alpha\beta}$ are positive semi-definite, so 
\be
0 < Z_{\alpha\beta} \leq \Zbar_{\alpha\beta} \leq\delta_{\alpha\beta}.
\label{eq:zbarabbounds}
\ee 
The covariance matrix of the theoretical autopredictions $T(f,\lambda)\equiv T[f]+\lambda_\alpha\beta_\alpha$,  Eq.~(\ref{eq:covTlamone},\ref{eq:covTlamfin}) then become 
\bea
{\Cov}[T(f,\lambda)] &=& X - S(C+S)^{-1}X-X(C+S)^{-1}S+\beta_\alpha\Zbar_{\alpha\beta}\beta_\beta^T \label{eq:covTlamf}\\
&=& C(C+S)^{-1}X(C+S)^{-1}C + S - S(C+S)^{-1}S,\label{eq:covTfitf}
\eea
the second expression being identical to the one we found in Sec.~\ref{subsec:autoprediction}. 

The same holds true of course for correlated predictions:
the shifts Eq.~(\ref{eq:shiftpred}) are now
\be
\label{eq:shiftpredf}
\delta \Ttil[f] = \betatil_\alpha\beta_\alpha^T(C+S)^{-1}(D-T[f_0]) = \Shat (C+S)^{-1}(D-T[f_0]),
\ee
where $\Shat = \betatil_\alpha\beta_\alpha^T$, while if $\Ttil(f,\lambda)=\Ttil[f]+\lambda_\alpha\betatil_\alpha$, Eq.~(\ref{eq:covTtillamone} becomes
\be
{\Cov}[\Ttil(f,\lambda)]
= \Xtil  - \Shat(C+S)^{-1}\Xhat^T-\Xhat(C+S)^{-1}\Shat^T+ \betatil_\alpha\Zbar_{\alpha\beta}\betatil_\beta^T \label{eq:covTtillamf},
\ee
where $\Stil = \betatil_\alpha\betatil_\alpha^T$, and
\bea
\Xtil&=& \Ttil_p(T_p^T(C+S)^{-1}T_q)^{-1}\Ttil_q^T,\label{eq:Xtildeff}\\
\Xhat&=& \Ttil_p(T_p^T(C+S)^{-1}T_q)^{-1}T_q^T.\label{eq:Xhatdeff}
\eea
Using Eq.~(\ref{eq:Zbardefab}), we can write the last term as
\bea
\betatil_\alpha\Zbar_{\alpha\beta}\betatil_\beta^T &=& \betatil_\alpha Z_{\alpha\beta}\betatil_\beta^T + \Shat(C+S)^{-1}X(C+S)^{-1}\Shat^T,\label{eq:ZbarStilab}\\
\betatil_\alpha Z_{\alpha\beta}\betatil_\beta^T &=&\Stil  - \Shat(C+S)^{-1}\Shat^T. 
\label{eq:ZStilab}
\eea
The final result thus again has exactly the same for as that found in Sec.~\ref{subsec:prediction}: once the nuisance parameters are all eliminated, the only changes are in the expressions for the covariances $X$, $\Xtil$ and $\Xhat$, generalizing the previous one parameter expressions to many parameters. 

\subsection{Fitting NNPDFs}
\label{subsec:fittingnnpdfs}

PDFs in an NNPDF fit are parametrized by a neural network, with a very large number of parameters. The fitting procedure differs from that using a fixed parametrization, since we want to avoid fitting noise. In practice this is achieved using a cross--validation procedure. It follows that when we fit to each data replica $D^{(r)}$, the neural net parameters, and thus the PDF replicas $f^{(r)}$,  are not precisely determined through exact minimization of the $\chi^2$, but rather include some random noise, which is responsible for the `functional uncertainty' inherent in the fit \cite{Ball:2014uwa}. It is not easy to describe this analytically: all we can say is that while all the general results in Sec.\ref{subsec:multiplenuisance} remain valid, relations such as Eq.~(\ref{eq:arep2f}) for the fitted parameters, and thus the subsequent results Eqs.~(\ref{eq:varaf},\ref{eq:Xdeffpq}), no longer hold. However we can still use Eq.~(\ref{eq:lambdabarfx}) to compute the expectation and covariance of the nuisance parameters, and obtain the same results Eqs.~(\ref{eq:varaf},\ref{eq:Zbardefab}), provided we define $T^{(r)}\equiv T[f^{(r)}]$, and $T^{(0)}\equiv \langle T^{(r)}\rangle$
\be
X\equiv\Cov[T[f]] = \langle (T^{(r)} - T^{(0)}) (T^{(r)} - T^{(0)})^T\rangle
\label{eq:XdefNN}
\ee
as averages over the PDF replicas. This matrix gives the PDF uncertainties (and correlations) for the observables $T[f]$, which includes both the experimental uncertainties in the data and the theoretical uncertainties in extracting the PDFs from the data.

Note that in an NNPDF fit  $X$ no longer satisfies the projective relation Eq.~(\ref{eq:XsqeqX}), and indeed $X(C+S)^{-1}$ no longer projects data replicas directly onto theory replicas as in Eq.~(\ref{eq:projection}).  It can confirm this for a given set of PDF replicas by computing the cross-covariance matrix
\be
Y \equiv {\rm Cov}[T,D]= \langle (T^{(r)} - T^{(0)}) (D^{(r)} - D)^T\rangle.
\label{eq:YdefNN}
\ee
For a fixed parametrization, we can use Eq.~(\ref{eq:projection}) and Eq.~(\ref{eq:repavD}) to show that then $Y=X=Y^T$. However it is easy to check by explicit computation that in an NNPDF fit $Y$ is generally considerably smaller than $X$: the fluctuations of the theory replicas are not very well correlated to the fluctuations of the data replicas due to the functional uncertainty. So although many of the eigenvalues of $X(C+S)^{-1}$ will still be zero (because $m<N$), the nonzero eigenvalues will differ from one, and many will be somewhat larger than one due to the functional uncertainty. This means that while Eq.~(\ref{eq:Zabbounds}) still holds, the upper bound on $\Zbar_{\alpha\beta}$, Eq.(\ref{eq:zbarabbounds}), does not: the covariance of the nuisance parameters can be larger than the prior when the functional uncertainty is large.

Note that the fact that $X$ is not invertible is not in any sense a technical limitation: the mapping of a global dataset into a set of PDFs cannot in principle be invertible (except possibly in certain special cases, such as data from a single process taken at a single scale \cite{Harland-Lang:2018bxd}), since it is impossible to recover the data solely from the PDFs. This is in part because the PDFs are only functions of $x$, while the data also depend on a scale: when we determine PDFs, all the data are effectively projected onto a common scale. But it is also because PDFs are by definition universal, i.e. process independent, so given a set of PDFs it is impossible in principle to say even which processes were used to determine them.

We can now derive general results for the expectation and covariance of autopredictions from the three matrices $C$, $S$ and $X$, following the procedure set out in Sec.~\ref{subsec:autoprediction}. The shifts in the autopredictions are given by a similar expression to Eq.~(\ref{eq:shiftmult}), 
\be
\label{eq:shiftNN}
\delta T[f]  = -S(C+S)^{-1}(T^{(0)}-D),
\ee
and will reduce the experimental $\chi^2$ as explained already in Sec.~\ref{subsec:puretheory}. The covariance matrix is still given by Eq.~(\ref{eq:covTfitf}):
\be
P\equiv{\Cov}[T(f,\lambda)] = C(C+S)^{-1}X(C+S)^{-1}C + (S - S(C+S)^{-1}S).\label{eq:PNN}
\ee
If the theory uncertainty $S$ is much smaller than the experimental uncertainty $C$, $P$ approaches the result
\be
P_{\rm con}= X+S;\label{eq:PconNN}
\ee
the fitting uncertainty and theoretical uncertainty can be combined in quadrature. So when the experimental uncertainties dominate there is almost complete decorrelation of the theoretical uncertainties, and the `conservative' prescription recommended in Ref.~\cite{AbdulKhalek:2019ihb} is a useful approximation. 

Returning to the more general correlated result Eq.~(\ref{eq:PNN}), both contributions to $P$ (which we may call the correlated PDF uncertainty and the correlated theory uncertainty) are also positive semi-definite, and combine in quadrature to give the total uncertainty. Moreover the correlated theory uncertainty bounded above by the corresponding uncorrelated theory uncertainty:
\be
0\leq S - S(C+S)^{-1}S = C(C+S)^{-1}S \leq S,\label{eq:Sbound}
\ee
It is tempting to also think that the correlated PDF uncertainty will also be bounded above by the uncorrelated PDF uncertainty $X$, because since $C$ is positive definite, and $S$ positive semi-definite, $C\leq C+S$, so $C(C+S)^{-1}\leq 1$, and $C(C+S)^{-1}X(C+S)^{-1}C \leq X$. This argument is wrong, however, and the correlated PDF uncertainty can sometimes exceed the uncorrelated. Writing
\be
C(C+S)^{-1}X(C+S)^{-1}C = X - S(C+S)^{-1}X- X(C+S)^{-1}S+S(C+S)^{-1}X(C+S)^{-1}S,
\label{eq:Xalgebra2}
\ee
in some circumstances the sum of the last three terms can be positive. For this reason it seems impossible to prove in general that $P\leq P_{\rm con}$, though in all practical applications we have tested so far this seems to be the case.

For genuine predictions, with theory uncertainties correlated to those in the fit, shifts are given by  Eq.~(\ref{eq:shiftpredf}), 
\be
\label{eq:shifttilNN}
\delta \Ttil[f]  = -\Shat(C+S)^{-1}(T^{(0)}-D),
\ee
while Eq.~(\ref{eq:covTtillamf}) is most usefully written in the form
\begin{eqnarray}
\Ptil&\equiv&{\Cov}[\Ttil(f,\lambda)]\nn\\
&=&\Xtil  - \Xhat(C+S)^{-1}\Shat^T - \Shat(C+S)^{-1}\Xhat^T+\Shat(C+S)^{-1}X(C+S)^{-1}\Shat^T\nn\\
&&\qquad\qquad\qquad\qquad + (\Stil  - \Shat(C+S)^{-1}\Shat^T) \label{eq:PtilNN},
\end{eqnarray}
where now besides the matrix $X$ Eq.~(\ref{eq:XdefNN}) we must now also evaluate 
\begin{eqnarray} 
\Xtil&\equiv&\Cov[\Ttil[f,\lambda]] = \langle (\Ttil^{(r)} -\Ttil^{(0)}) (\Ttil^{(r)} - \Ttil^{(0)})^T\rangle\label{eq:XtildefNN},\\
\Xhat&\equiv&\Cov[\Ttil[f,\lambda],T[f,\lambda]] = \langle (\Ttil^{(r)} -\Ttil^{(0)}) (T^{(r)} - T^{(0)})^T\rangle\label{eq:XhatdefNN}.
\end{eqnarray}
Again the covariance Eq.~(\ref{eq:PtilNN}) separates into the sum in quadrature of a correlated PDF uncertainty (the first line) and a correlated theory uncertainty (the second line), When the cross-covariance $\Shat$ is very small, we obtain the conservative result
\be
\Ptil_{\rm con} =  \Xtil+\Stil,  \label{eq:PtilconNN}
\ee 
proposed in Ref.~\cite{AbdulKhalek:2019ihb}. This will typically be the case for predictions of new processes where the dominant MHOU is in the hard cross-section. However for processes already included in the fit, the situation is more complex, since $\Stil$ and $\Shat$ may be large even if $S$ is small.

Note that since the inclusion of MHOU in the PDF determination leads to only a small increase in the uncertainties of the PDFs~\cite{AbdulKhalek:2019ihb}, the conservative results Eq.~(\ref{eq:PconNN},\ref{eq:PtilconNN}) give uncertainties very close to the conventional prescription, in which the PDF uncertainty (without MHOU) is combined in quadrature with the MHOU in the prediction. This will be particularly true when the MHOU in the prediction is larger than the PDF uncertainty.

\section{Numerical Results}
\label{sec:numeric}

\begin{table}[b!]
  \centering
  \renewcommand*{\arraystretch}{1.3}
  \begin{tabular}{|c|c|}
    \hline
    Process Type  & Datasets \\
    \hline
    DIS NC  &   SLAC, BCDMS, NMC, HERA NC \\
    DIS CC  &   NuTeV, CHORUS, HERA CC \\
    DY  & CDF, D0, ATLAS, CMS, LHCb (dileptons, $W$ and $Z$ diff xsecs) \\
    JET  & ATLAS, CMS inclusive jets \\
    TOP  & ATLAS, CMS total \& diff xsecs \\
    \hline
  \end{tabular}
  \caption{\label{tab:expclassification}
   Classification of  datasets into  process types.
  }
\end{table}

In Sec.~\ref{subsec:fittingnnpdfs} we saw that in a realistic global PDF fit that we still use the same analytic expressions Eqs.~(\ref{eq:shiftNN},\ref{eq:PNN},\ref{eq:shifttilNN},\ref{eq:PtilNN}) for the shifts and reduction in uncertainties induced by the correlations between the theoretical uncertainties in fit and prediction as we would use in a fit with a fixed parametrization Sec.~\ref{subsec:pdfexactfit}. This is despite the fact that PDFs are smooth functions, which cannot be determined uniquely from a finite set of discrete data, but necessarily have an additional `functional uncertainty', so the PDF parameters are not fixed uniquely by the fit. All that is necessary is to evaluate the matrices $X$, $\Xtil$ and $\Xhat$ Eqs.~(\ref{eq:XdefNN},\ref{eq:XtildefNN},\ref{eq:XhatdefNN}) as ensemble averages over the PDF replicas determined in the fit. In this section we will compute these matrices in a realistic global PDF fit with theory uncertainties, and use them to evaluate autopredictions and genuine predictions including the effect of the correlated theoretical uncertainties.

\begin{figure}[h!]
    \begin{center}
    \makebox{\includegraphics[width=0.48\columnwidth]{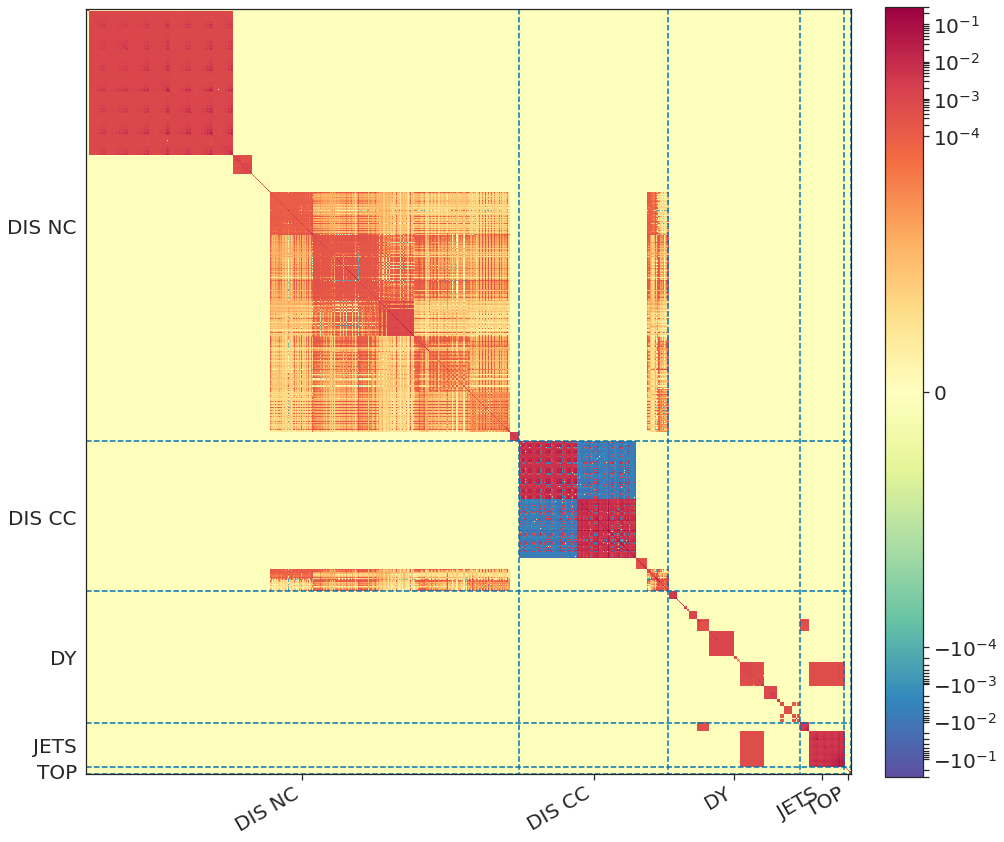}}
\makebox{\includegraphics[width=0.48\columnwidth]{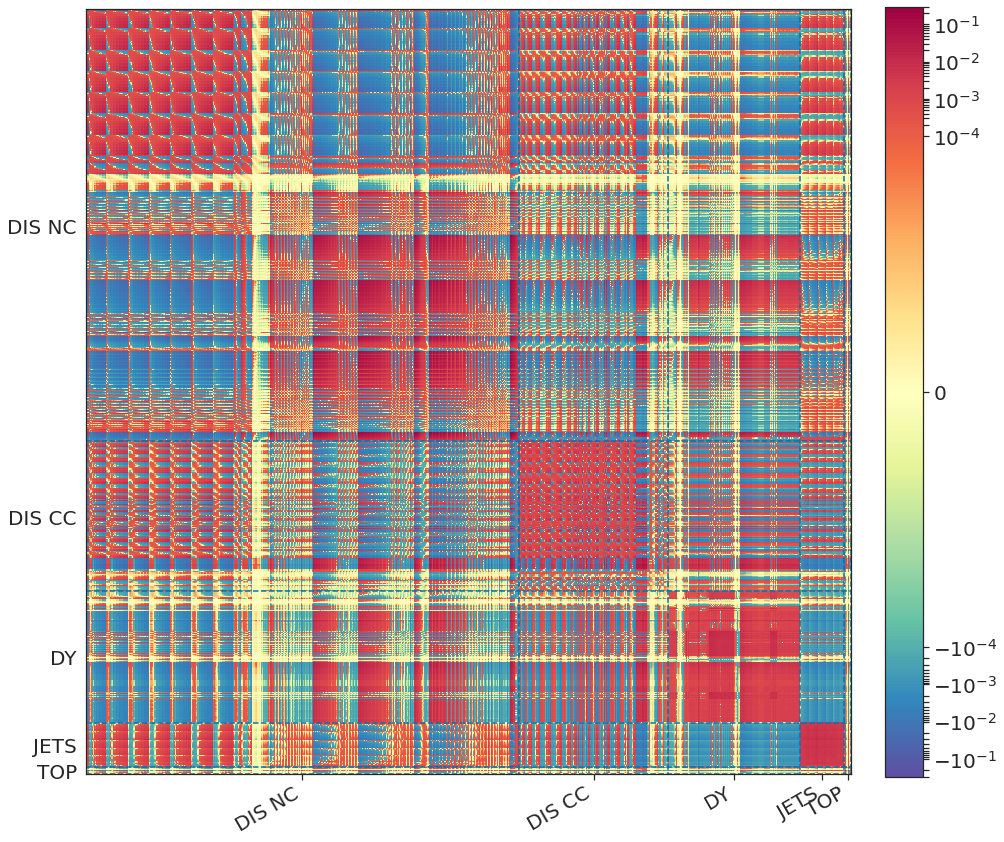}}
    \end{center}
  \vspace{-0.55cm}
  \caption{The experimental covariance matrix, $C_{ij}$, normalized to the theoretical predictions $T^{(0)}_i$  (left), and the corresponding theory covariance matrix for MHOU, $S_{ij}$ (right). The datasets are arranged in the order given in Fig.~\ref{fig:chi2auto} below: so SLAC data are in the top left corner, and LHC top data in the lower right corner.}
  \label{fig:CnS}
\end{figure}

We will carry out these studies in the context of the NNPDF3.1 NLO global fit with MHOU presented in Ref.~\cite{AbdulKhalek:2019ihb}.
This in turn employed the same experimental data and theory calculations in NNPDF3.1~\cite{Ball:2017nwa} with two minor differences:
the value of the lower kinematic
cut was increased from
$Q_{\rm min}^2=2.69$~GeV$^2$ to
$13.96$~GeV$^2$, and the HERA $F_2^b$, fixed-target Drell-Yan cross-sections, and some LHC inclusive jet data were removed, for technical
reasons. This left a total of $N_{\rm dat}=2819$ data points. The complete list of data included in the fit may be found in Tab.~6.3 of Ref.~\cite{AbdulKhalek:2019ihb}. These data were divided into five classes, depending on the type of process involved,
as summarized in Tab.~\ref{tab:expclassification}. The MHOU covariance matrix $S_{ij}$ was constructed using renormalization and factorization scale variations, by a factor of two either side. The factorization scale variations (estimating the MHOU in the NLO parton evolution) are correlated across all processes, but the renormalization scale variations (estimating the MHOU in the NLO hard cross-sections peculiar to each process) while correlated within data belonging to the same process, are uncorrelated between different processes. These variations were then combined to give  $S_{ij}$ using the 9pt scheme, as explained in Ref.~\cite{AbdulKhalek:2019ihb}. The matrices $C_{ij}$ and $S_{ij}$ computed in Ref.~\cite{AbdulKhalek:2019ihb} are reproduced in Fig.~\ref{fig:CnS} as heat maps.

\subsection{Covariance of PDF uncertainties $X$}
\label{subsec:XnY}

\begin{figure}[h!]
    \begin{center}
    \makebox{\includegraphics[width=0.48\columnwidth]{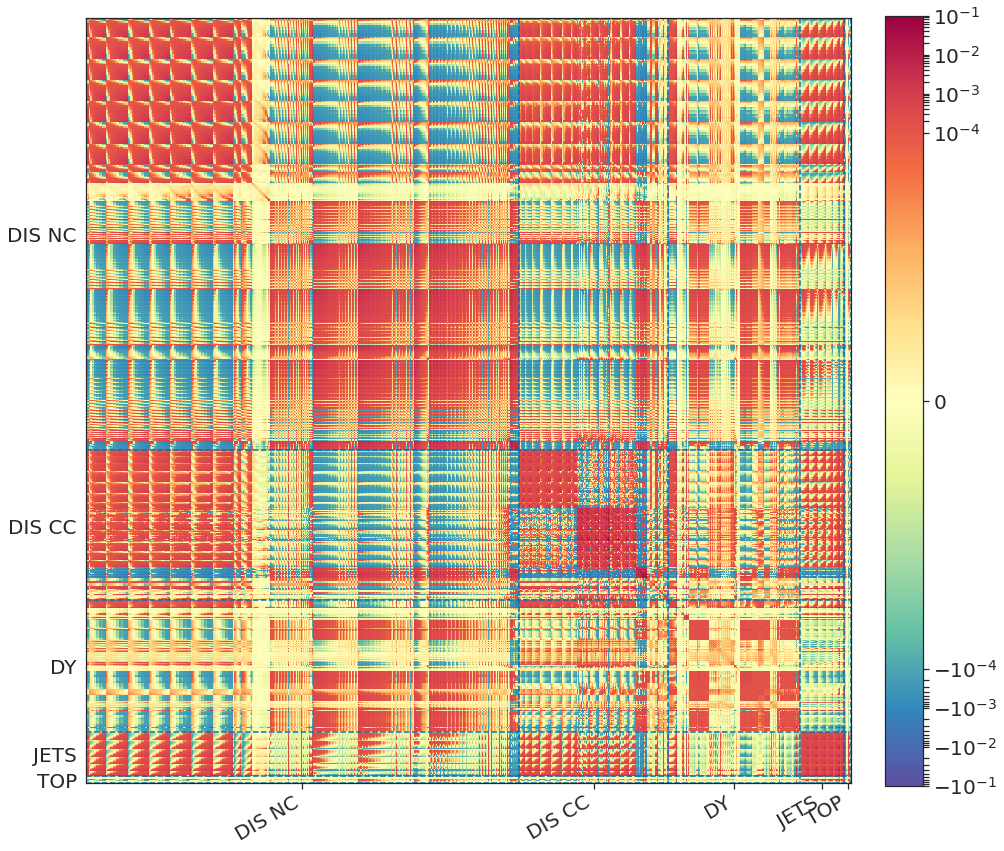}}
\makebox{\includegraphics[width=0.48\columnwidth]{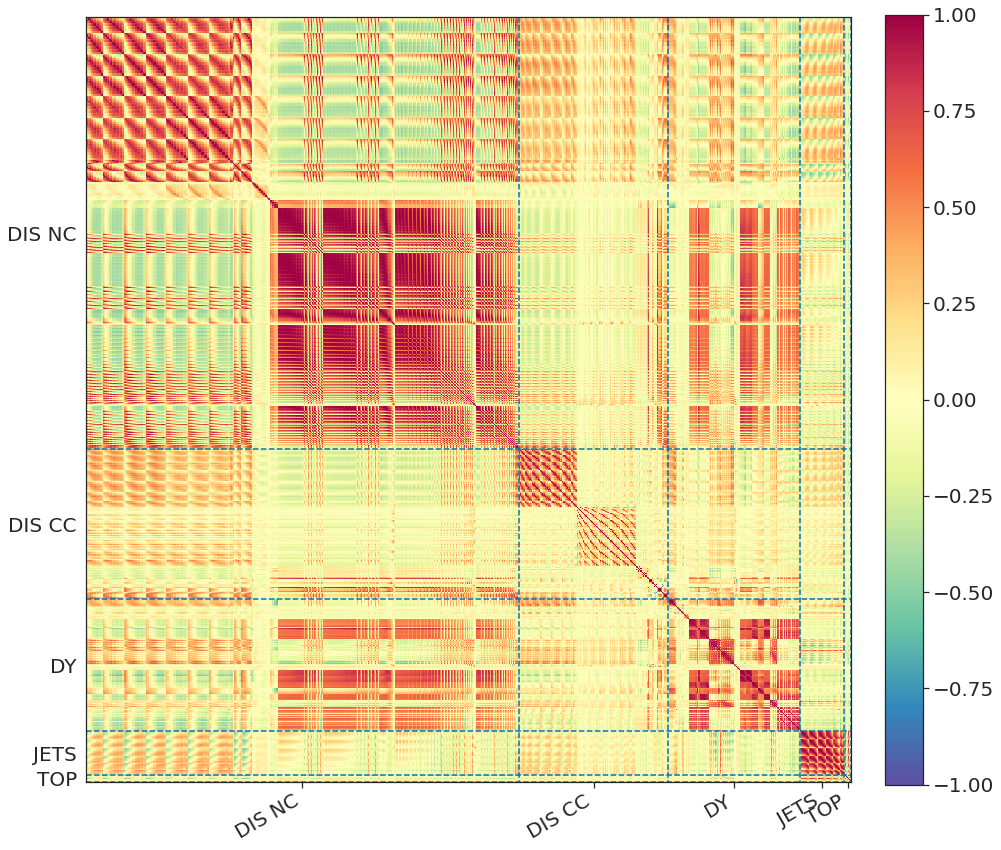}}
    \end{center}
  \vspace{-0.55cm}
  \caption{The covariance matrix of PDF uncertainties, $X_{ij}$, normalized to the theoretical predictions $T^{(0)}_i$  (left), and the corresponding correlation matrix $X_{ij}/\sqrt{X_{ii}X_{jj}}$ (right). The datasets are arranged in the order given in Fig.~\ref{fig:chi2auto} below: so SLAC data are in the top left corner, and LHC top data in the lower right corner.}
  \label{fig:X}
\end{figure}

We begin by computing the covariance matrix $X_{ij}$, Eq.~(\ref{eq:XdefNN}), shown in Fig.~\ref{fig:X} as a heat map alongside the corresponding correlation matrix.  
It can be seen that the off-diagonal elements of $X_{ij}$ are almost as large as the diagonal elements: this is confirmed by examination of the correlation matrix. This is because theoretical predictions are often very strongly correlated, not only for nearby bins within the same experiment, but also for different processes at nearby scales, due primarily to the smoothness of the underlying PDFs, both in $x$ and in $Q^2$, but also due to the highly correlated theoretical uncertainties included in the fit. 

 \begin{figure}[t!]
    \begin{center}
    \makebox{\includegraphics[width=0.99\columnwidth]{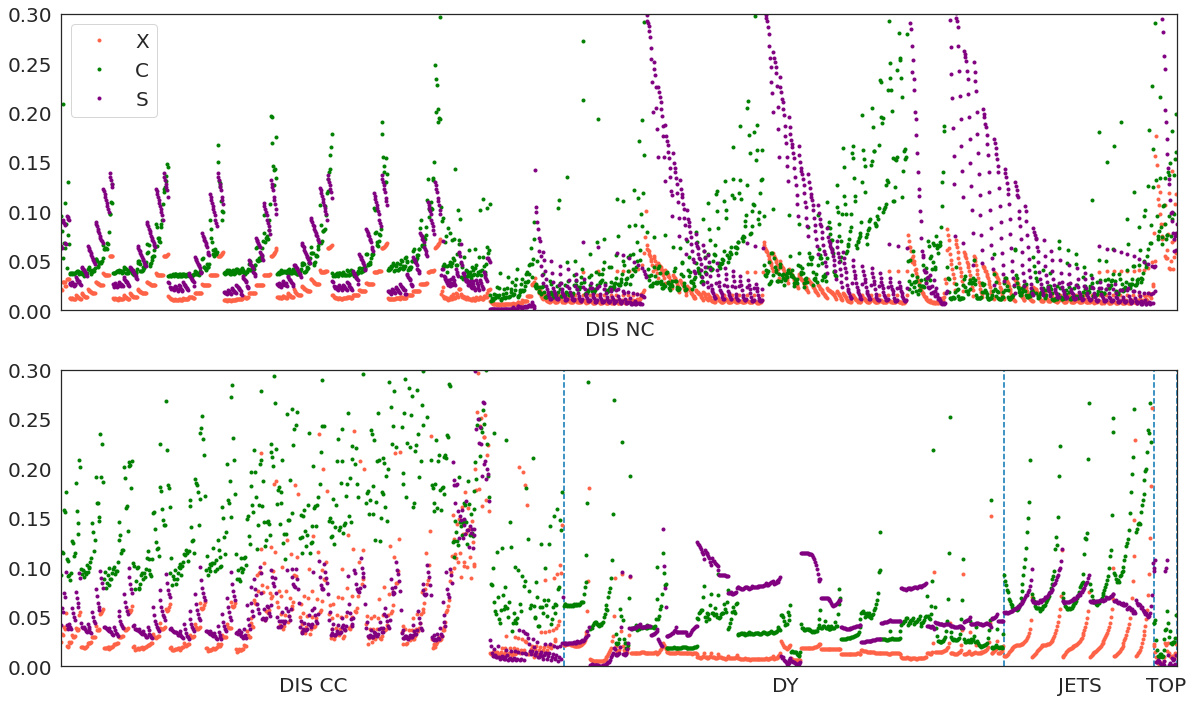}}
    \end{center}
  \vspace{-0.55cm}
  \caption{The square root of the diagonal elements of the matrices $X$ (in orange), $C$ (in green) and $S$ (in purple) normalized to the theoretical predictions $T^{(0)}_i$, with those for $C$ and $S$ the same as in Ref.~\cite{AbdulKhalek:2019ihb}. The datasets are arranged in the order given in Fig.~\ref{fig:chi2auto} below.}
  \label{fig:CXS}
\end{figure}


We compare the PDF uncertainties to the experimental and theoretical uncertainties by looking at the per-point uncertainty (Fig.~\ref{fig:CXS}). Recall (Eqn.~(\ref{eq:repavD})) that $C+S$ is the covariance of the data replicas to which the PDF replicas are fitted. It can be seen that at NLO the relative size of the experimental uncertainties $C_{ii}$ and the theoretical uncertainties $S_{ii}$ varies considerably between different datasets: for the fixed target DIS data $S_{ii}$ is generally below $C_{ii}$, except at large $x$, whereas for the HERA NC data $S_{ii}$ is much less than $C_{ii}$ at large $x$, but the other way around at small $x$ where the theoretical uncertainty dominates. For CHORUS, the experimental uncertainty also dominates, while for most DY datasets the theoretical uncertainty dominates. In contrast the PDF uncertainties $X_{ii}$ are generally less than either $C_{ii}$ or $S_{ii}$, because the combination of data within given datasets and information from other datasets in the fit conspire to reduce the uncertainty. This is especially evident for DIS CC, DY and JETS. However for some data sets, particularly cross-section ratios with very small theory uncertainty (such as NMC $d/p$, the asymmetries, and the differential top data), $X_{ii}$ lies above $S_{ii}$, though still below $C_{ii}$. 

\subsection{Nuisance Parameters}
\label{subsec:nuisance2}

  \begin{figure}[t!]
    \begin{center}
    \makebox{\includegraphics[width=0.7\columnwidth]{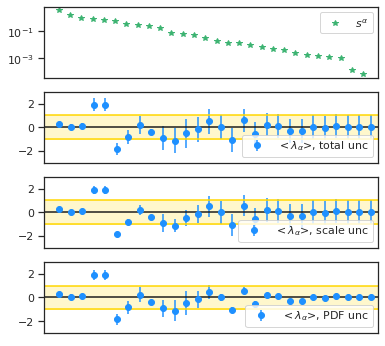}}
    \end{center}
  \vspace{-0.55cm}
  \caption{The $28$ positive eigenvalues $s^\alpha$ of the theory uncertainty matrix $S_{ij}$ (above), shown in descending order, and 28 nuisance parameters $\lambda_\alpha$ corresponding to the $28$ eigenvectors $\beta_\alpha$ (below), as given by Eq.~(\ref{eq:Elambdaf}).The uncertainties in the nuisance parameters are shown in total (square roots of the diagonal entries of Eq.~(\ref{eq:Zbardefab}), and broken down into the contribution from scale uncertainties alone (square roots of the diagonal entries of Eq.~(\ref{eq:Zdefab})  and from PDF uncertainties (square roots of the diagonal entries of the last term in Eq.~(\ref{eq:Zbardefab}). The yellow bands highlight the region between $\pm 1$.}
  \label{fig:nuisancediag}
    \begin{center}
    \makebox{\includegraphics[width=0.6\columnwidth]{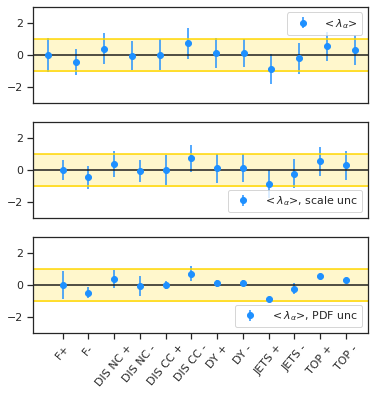}}
    \end{center}
  \vspace{-0.55cm}
  \caption{Nuisance parameters $\lambda$ for directions in the space of scale variations corresponding to up/down changes in factorization scale, and in renormalization scale for the five types of processes in the determination of the $9$-pt theory covariance matrix for MHOU. The uncertainties in the nuisance parameters are shown in total, and broken down into the contribution from scale uncertainties alone and from PDF uncertainties, just as in Fig.~\ref{fig:nuisancediag}. The yellow bands highlight the region between $\pm 1$.}
  \label{fig:nuisancephys}
\end{figure}

Having computed $X$, we next calculate the nuisance parameters $\lambda_\alpha$ of the theory covariance matrix $S_{ij}$ for the MHOU in the  NNPDF3.1 NLO global fit. The posterior distributions of these nuisance parameters give us information on which directions of MHOU are constrained by the fit. We showed in Ref.~\cite{AbdulKhalek:2019ihb} that when there are five different
processes, there are 28 nonzero eigenvalues. Thus we have 28 nuisance parameters, in one-to-one correspondence with the 28 eigenvectors with nonzero eigenvalues. The eigenvalues are shown in descending order in Fig.~\ref{fig:nuisancediag}, with the nuisance parameters below them. The expectation values of the nuisance parameters are computed using  Eq.~(\ref{eq:Elambdaf}), and their uncertainties using Eq.~(\ref{eq:Zbardefab}). The nuisance parameters are normalized so that their prior is a unit gaussian centred on zero, as in Eq.~(\ref{eq:priorf}). It can be seen that after fitting, the uncertainty in the nuisance parameters associated with the largest nine or so eigenvalues has been substantially reduced from one, indicating that exposure to the data has reduced the MHOUs. For those corresponding to the smaller eigenvalues there is very little reduction, showing that the data do not much constrain these directions in the space of MHOU. The central values for the three largest eigenvalue nuisance parameters remain close to zero within uncertainties, showing that the prior choices (mainly overall normalizations) were reasonable, while the next three or four show significant deviations from zero: for these the data seem to carry significant information about the MHOU. For the remaining (smaller) eigenvalues the central values of nuisance parameters are all consistent with zero, and clearly for the very small ones the data have no effect at all, the posterior distributions being the same as the prior. This shows that only the eigenvectors corresponding to the larger eigenvalues are actually relevant for the PDF determination: the remainder correspond to such small changes in theoretical uncertainty that the fit ignores them.

We can understand these features better by separating out the two contributions to the total uncertainty in the nuisance parameters: that due purely to the impact of the fit of a given replica on the MHOU (given by  Eq.~(\ref{eq:Zdefab})), and that due to the additional PDF uncertainty when the fits to all the replicas are averaged over PDF replicas, given by the last term in Eq.~(\ref{eq:Zbardefab}), also shown in  Fig.~\ref{fig:nuisancediag}. We see that when fitting a single replica, the uncertainties in the nuisance parameters corresponding to the larger eigenvalues are indeed very substantially reduced: the MHOU along the eigenvectors corresponding to these larger eigenvalues is learnt in the fit to the data, just as we saw in the simple models in Sec.~\ref{subsec:puretheory}. Again, very little information is retained about the smaller eigenvalues. The uncertainty contributed by averaging over the PDF replicas is also small for the largest eigenvalue nuisance parameters, but becomes the dominant contribution after the first three. For the smallest it is very small again: for these the data have no effect. 

We can learn a little more about which MHOUs are learnt most by choosing different directions for the shift vectors $\beta_\alpha$  in Eq.~(\ref{eq:PTDlf}) than for the eigenvectors of $S_{ij}$. Specifically, we can choose the $\beta_\alpha$ to correspond to factorization scale variations (up or down), or renormalization scale variations (up or down, but now separately for each process). The results are shown in  Fig.~\ref{fig:nuisancephys}, where we again show the total uncertainty, scale uncertainty, and PDF uncertainty. The central values fluctuate about zero, but all remain within the band $\pm 1$, showing that the effect of fitting the experimental data on the nuisance parameters is rather mild: this is reassuring, as it confirms the choice of central scales used to make the predictions, and the choice of the range of the scale variations (implicit in the choice of the prior for the nuisance parameters, Eq.~(\ref{eq:priorf})). We also see that the uncertainties in the nuisance parameters corresponding to factorization scale variations (estimating the MHOUs in parton evolution), are reduced the most when fitting to any given replica, as we would expect since MHOUs in parton evolution are common to all data included in the fit. A little is also learnt about the renormalization scales for DIS. However, the PDF uncertainties partially wash out these effects. This suggests that the significant shifts in the nuisance parameters seen in 
Fig.~\ref{fig:nuisancediag} are due to global tensions between different processes, rather than problems with the choice of scales for particular processes.

Already we see that information from the data in the fit significantly updates the priors for the nuisance parameter distribution. From this it is likely that there will be an effect at the level of autopredictions, which is the subject of the next section.

\subsection{Autopredictions}
\label{subsec:autopredictions}

We now present results for the `autopredictions'. As we explained already in Sec~\ref{subsec:puretheory}, these are the theoretical predictions we make for all the datasets included in the PDF fit, including theoretical uncertainties, after the fitting of the PDFs (with these same theoretical uncertainties). They can thus be thought of as `postdictions', or predictions for the results of experiments run in exactly the same way, with the same equipment, as the original experiment, but taking account of the original global dataset. They thus form an ideal theoretical laboratory for testing the extent of the decorrelation between the theoretical uncertainties in the PDF fit, and those in the (auto)predictions.
\begin{figure}[t!]
    \begin{center}
    \makebox{\includegraphics[width=0.95\columnwidth]{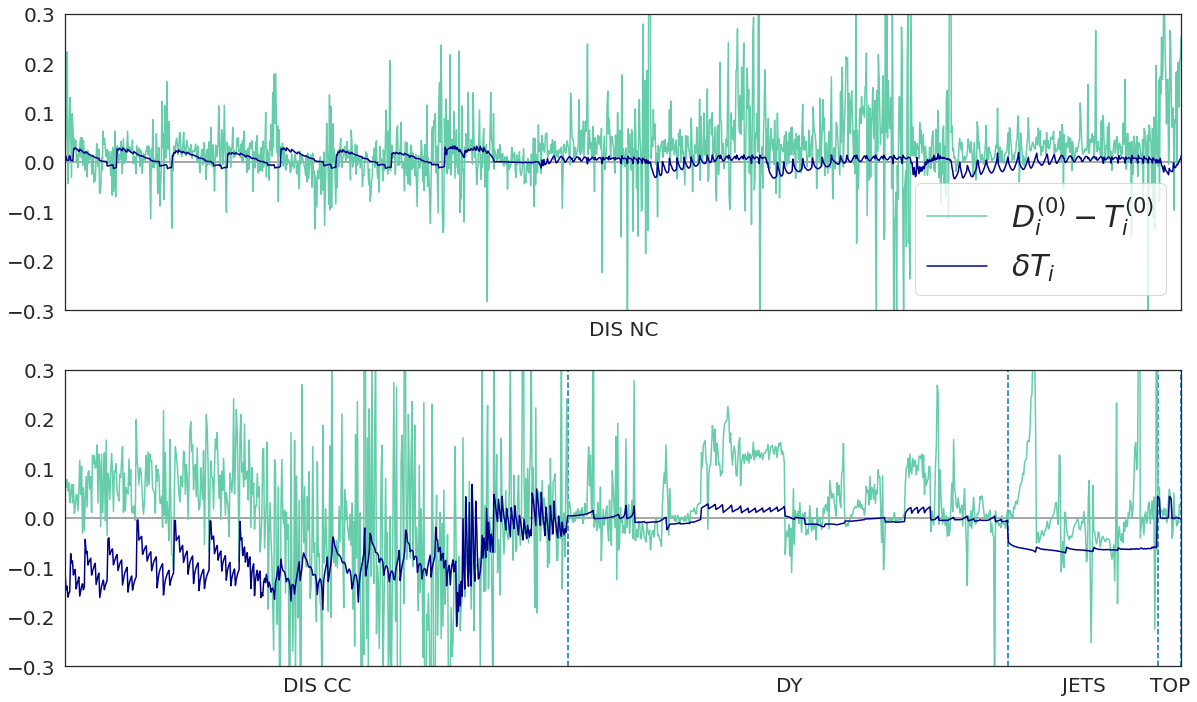}}
    \end{center}
  \vspace{-0.55cm}
  \caption{The shifts $\delta T_i$, Eq.~(\ref{eq:shiftNN}) (in blue) compared to the differences between theory and data, $D_i-T^{(0)}_i$ (in green), both normalized to $T^{(0)}_i$.} 
  \label{fig:shifts}
    \begin{center}
    \makebox{\includegraphics[width=0.99\columnwidth]{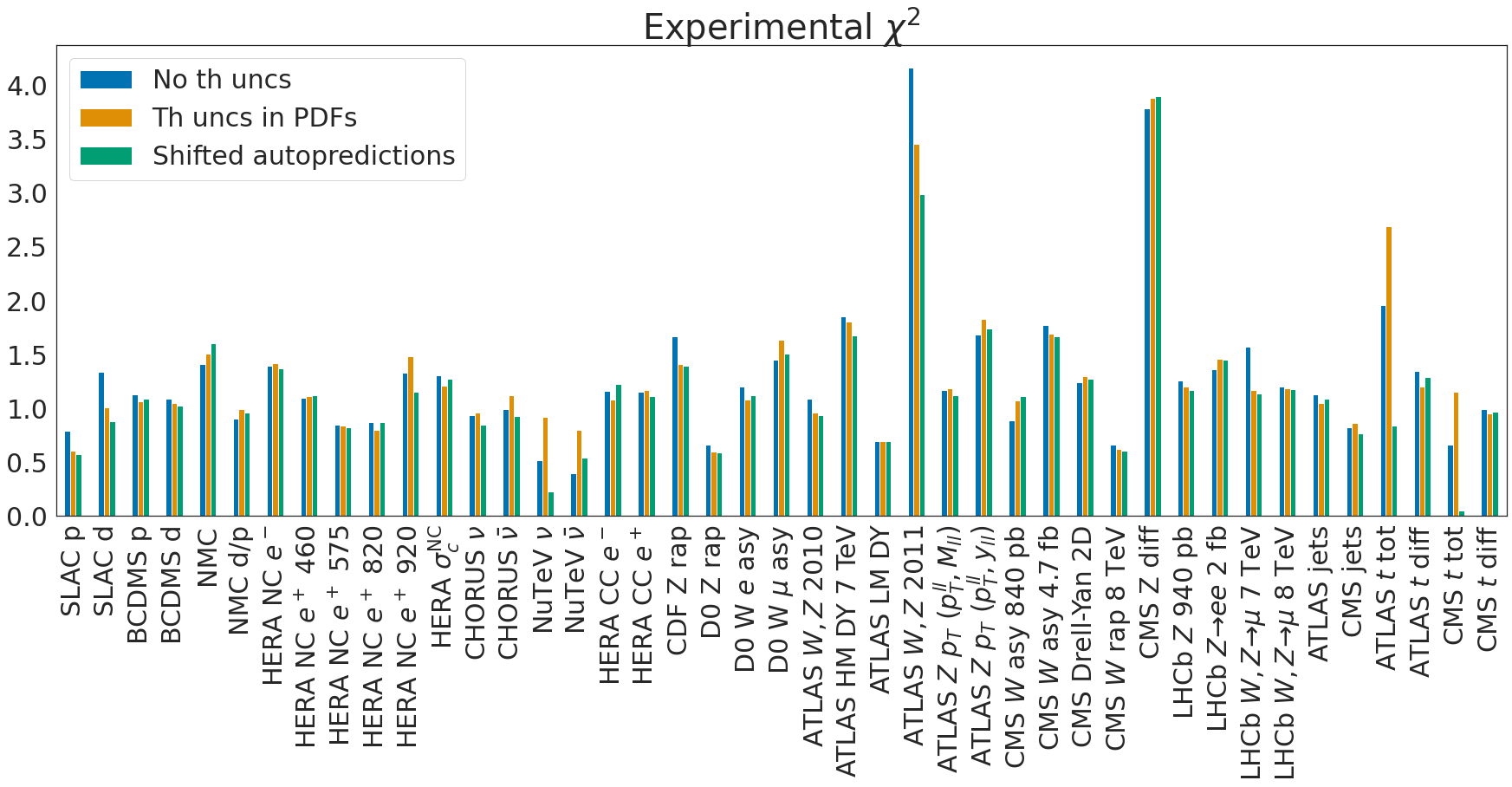}}
    \end{center}
  \vspace{-0.55cm}
  \caption{The experimental $\chi^2$ for each data set, comparing the original result of the NLO fit with no theory uncertainties to the fit with theory uncertainties, and then including the correlated shift in the autopredictions.}
  \label{fig:chi2auto}
\end{figure}

We begin by computing the shifts $\delta T_i$, Eq.~(\ref{eq:shiftNN}), in the autopredictions, due to the correlation in theoretical uncertainty: these are shown in Fig.~\ref{fig:shifts}, normalized to the orginal theoretical prediction $T^{(0)}_i$. We also show for comparison the differences $D_i - T^{(0)}_i $. It can be seen from the plot that these shifts are generally much smaller than the difference between data and theory, particularly for DIS NC and DY. However for some datasets (in particular CHORUS and inclusive jets), there seems to be an overall shift in central value of the same order as the difference between experiment and theory. However it is difficult to draw any further conclusions from these observations, since the shifts are very correlated within datasets.

\begin{table}[b!]
\centering
\begin{tabular}{|l||ccccc|c|}
\hline
                   & \textbf{DIS NC} & \textbf{DIS CC} & \textbf{DY} & \textbf{JETS} & \textbf{TOP} & \textbf{Total} \\
                   \hline
No th uncs         & 1.13            & 0.98            & 1.56        & 0.88         & 1.20         & 1.17           \\
Uncorr th uncs     & 1.15            & 1.06            & 1.53        & 0.90         & 1.27         & 1.19           \\
Correlated th uncs & 1.09            & 0.91           & 1.47        & 0.83         & 0.97        & 1.10    \\
\hline
\end{tabular}
\vspace{1cm}
\caption{The experimental $\chi^2$ per data point for each process, comparing the original result of the NLO fit with no theory uncertainties to the fit with theory uncertainties, and then including the shift in the autopredictions.}
  \label{tab:chisq}
\end{table}

\begin{figure}[t!]
    \begin{center}
    \makebox{\includegraphics[width=0.48\columnwidth]{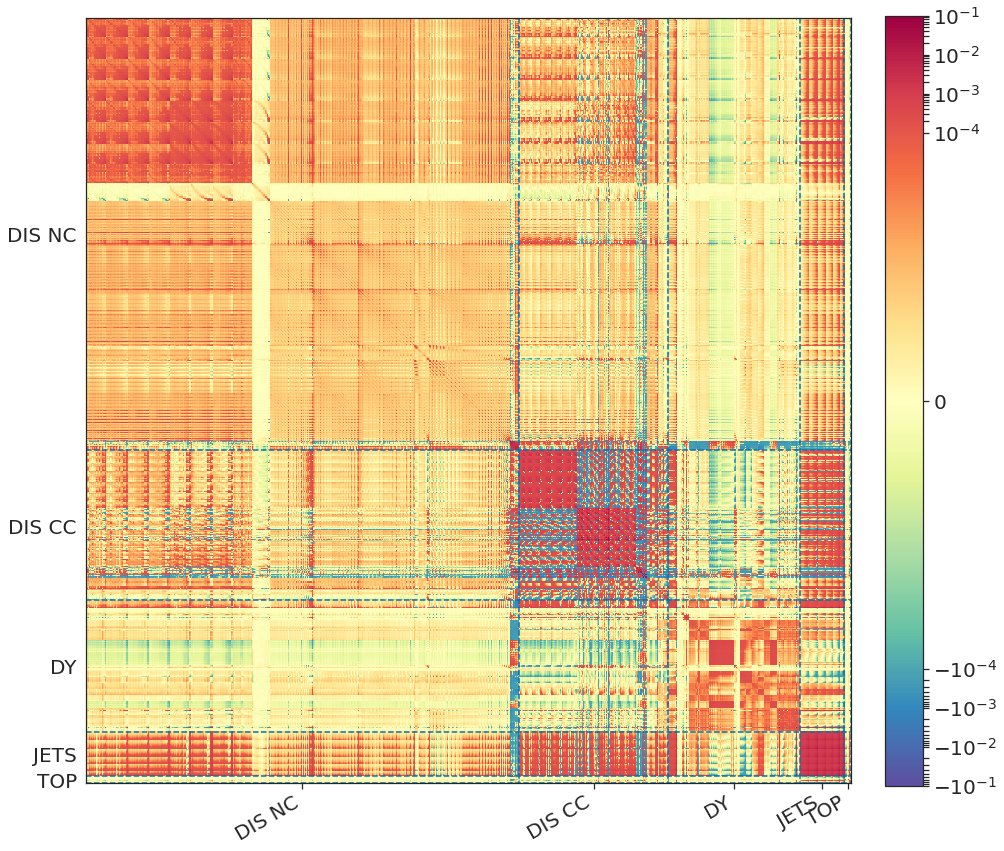}}
    \makebox{\includegraphics[width=0.48\columnwidth]{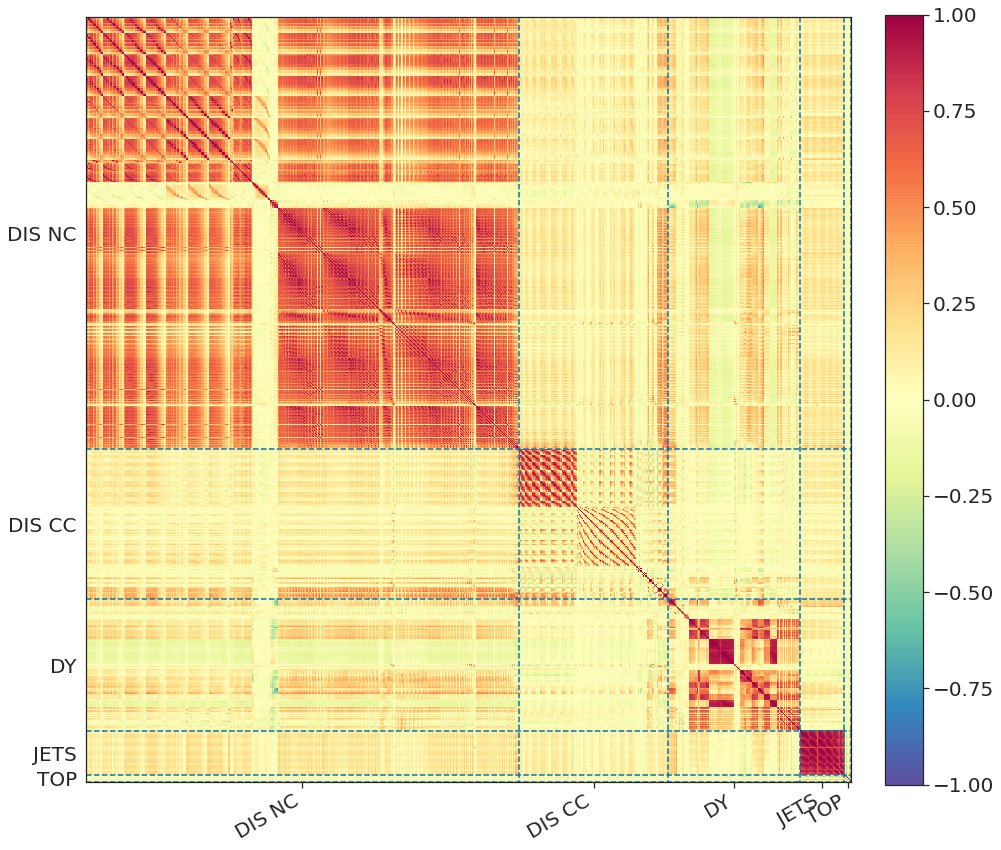}}  
  \end{center}
  \vspace{-0.55cm}
  \caption{The autoprediction covariance matrix $P_{ij}$ Eq.~(\ref{eq:PNN}), normalized to the theoretical predictions $T^{(0)}_i$ (left), and the corresponding corrrelation matrix $P_{ij}/\sqrt{P_{ii}P_{jj}}$ (right).}
  \label{fig:P}
    \begin{center}
    \makebox{\includegraphics[width=0.99\columnwidth]{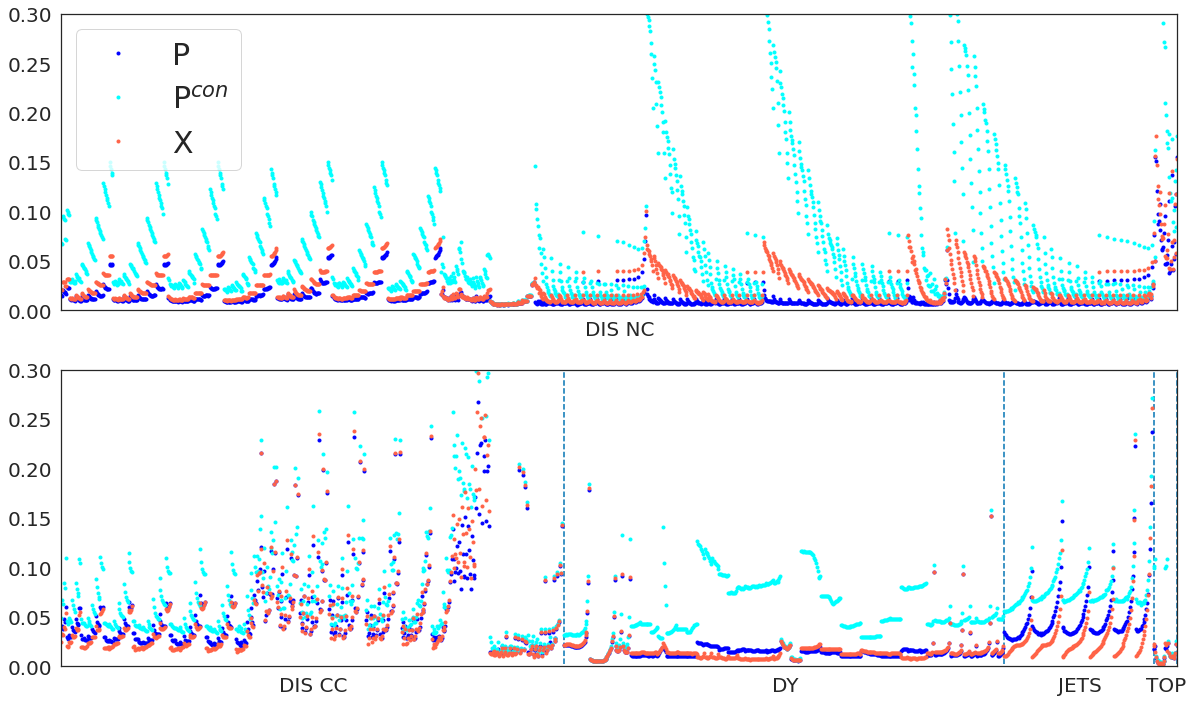}}
    \end{center}
  \vspace{-0.55cm}
  \caption{The percentage uncertainties of the autopredictions $\sqrt{P_{ii}}$ Eq.~(\ref{eq:PNN}) (blue) compared to the PDF uncertainty $\sqrt{X_{ii}}$ (orange),  and the conservative result, $\sqrt{P^{\rm con}_{ii}}$ Eq.~(\ref{eq:PconNN}) (cyan), all normalised to the theoretical predictions $T^{(0)}_i$.}
  \label{fig:Pdiag}
\end{figure}

To see whether the shifts actually improve the autopredictions, in Fig.~\ref{fig:chi2auto} we thus show the experimental $\chi^2$ for the original data, computed using the autopredictions for the fit with no theory uncertainties, those when the theory uncertainties are included in the fit, and then when the autoprediction includes the shift.  Needless to say all three results are generally very close, and including the theory uncertainties in the fit has mixed results, some predictions getting better, but at the expense of others getting worse, since the main effect of the theory uncertainties is to rebalance the datasets in the fit \cite{AbdulKhalek:2019ihb} . Nevertheless when the correlated shifts are included, the fit to most datasets improves, sometimes quite substantially, just as anticipated in the very simple `pure theory' model in Sec.~\ref{subsec:puretheory}, and confirmed for the more realistic models in Sec.~\ref{subsec:autoprediction} and Sec.~\ref{subsec:pdfexactfit}. The numbers broken down by process are shown in Tab.~\ref{tab:chisq}. When including theory uncertainties the total $\chi^2$ increases just a little, from $1.17$ to $1.19$, but when the correlated shift is added to the theoretical predictions, we see a significant improvement to $1.10$. This improvement is seen across all the processes. 

Although these autopredictions are in some sense artificial --- in practice experiments are never repeated using exactly the same equipment and settings --- the implications of this exercise for the learning of theoretical uncertainties are nevertheless rather general. This is because in a global fit of the size of that performed here, with 2819 data points from 35 datasets, involving five different processes, removing any one of the smaller datasets has very little impact on the PDFs, and removing any dataset has the effect of increasing PDF uncertainties, while theoretical uncertainties for the remaining data remain unchanged. Consequently if we were to perform the PDF fit without a given dataset, and repeat the analysis, so that the autoprediction becomes a genuine prediction (or more properly `postdiction'), the result for this genuine prediction would be very close to the autoprediction. So we expect the shifts in central values that we see in the autopredictions to be also give improvements in such predictions: the shifts should improve the accuracy of the such predictions. 

\begin{figure}[t!]
    \begin{center}
    \makebox{\includegraphics[width=0.9\columnwidth]{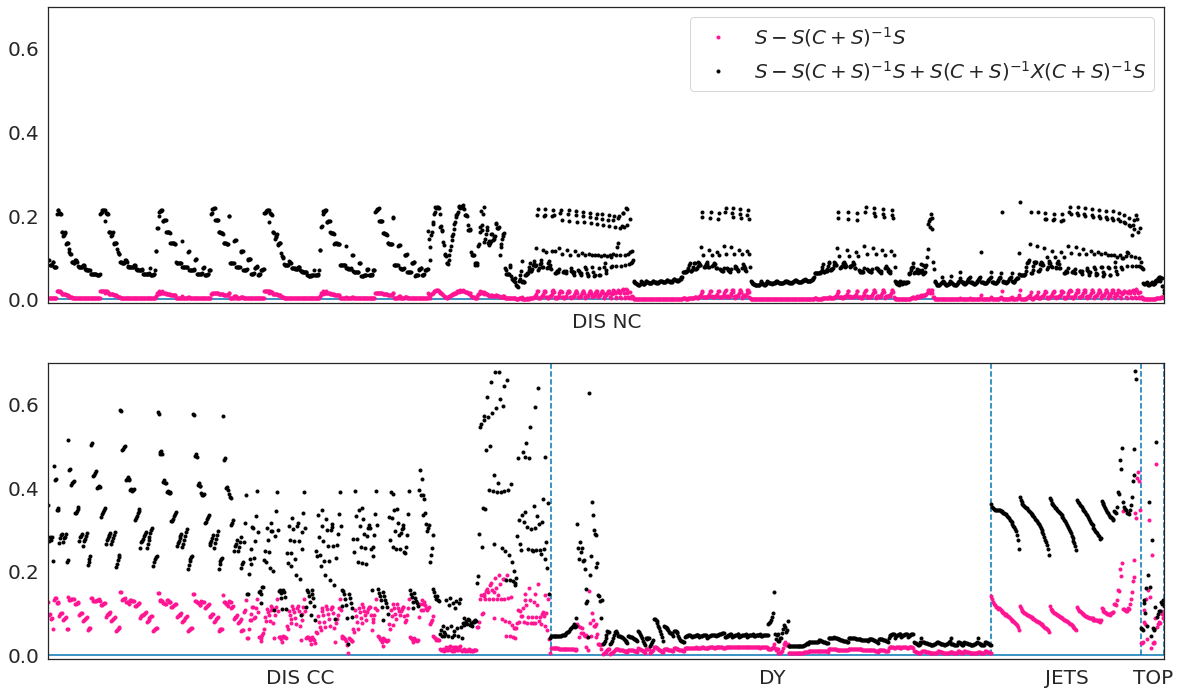}}
    \end{center}
  \vspace{-0.55cm}
  \caption{The contributions to the diagonal elements of the correlated theory uncertainty normalised to diagonal elements of S: $(S-S(C+S)^{-1}S)_{ii}/S_{ii}$ (pink), and  $(S-S(C+S)^{-1}S+S(C+S)^{-1}X(C+S)^{-1}S )_{ii}/S_{ii}$ (black).}
  \label{fig:Scpts}
    \begin{center}
    \makebox{\includegraphics[width=0.9\columnwidth]{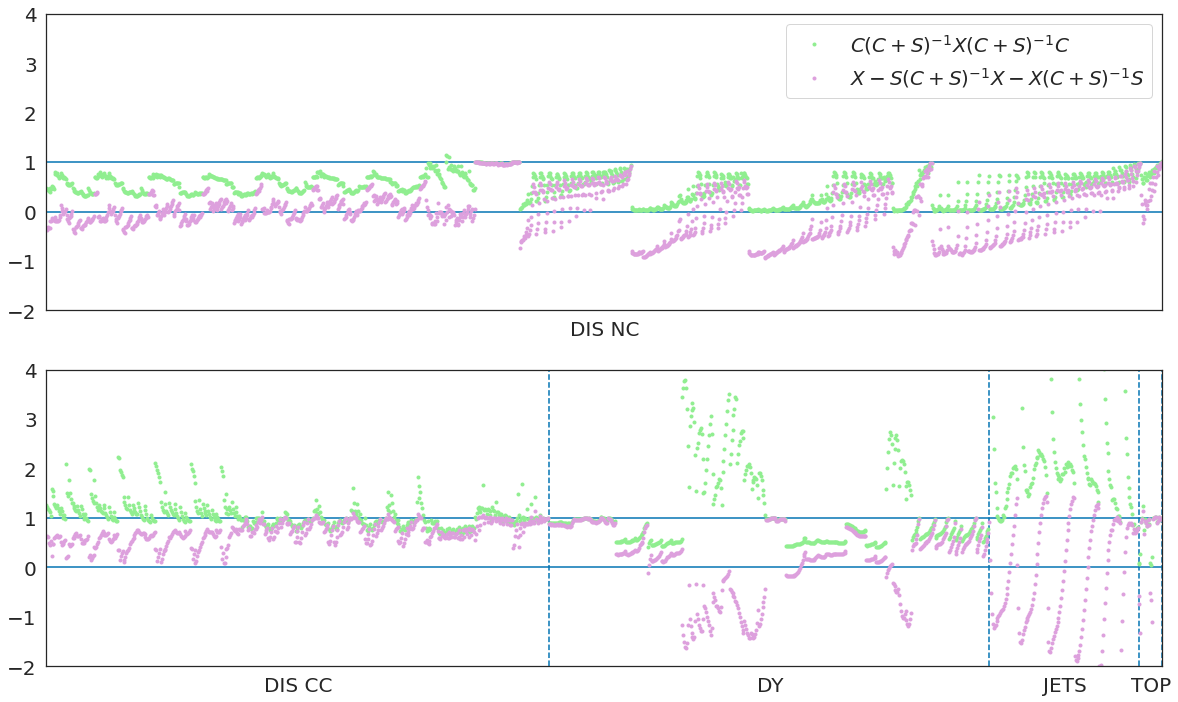}}
    \end{center}
  \vspace{-0.55cm}
  \caption{The contributions to the diagonal elements of the correlated PDF uncertainty normalised to diagonal elements of X: $(X-S(C+S)^{-1}X-X(C+S)^{-1}S)_{ii}/X_{ii}$ (lilac), and  $(C(C+S)^{-1}X(C+S)^{-1}C )_{ii}/X_{ii}$  (see Eq.~(\ref{eq:Xalgebra2}) (green).}
  \label{fig:Xcpts}
\end{figure}

To see whether we can also increase the precision, we consider the uncertainties in the autopredictions. In Fig.~\ref{fig:P} we show the full covariance matrix $P_{ij}$, Eq.~(\ref{eq:PNN}), again normalized to the theoretical predictions, and corresponding correlation matrix. The matrix $P_{ij}$ is the sum of the PDF uncertainty (derived from data uncertainties and theoretical uncertainties combined) and the theoretical uncertainty in the autoprediction, each reduced in size to account for the learning of the theoretical uncertainties and the correlation between the two sources of theoretical uncertainty. As might be expected there are very large correlations in the autopredictions within datasets, particularly for nearby kinematic points, but there are also smaller correlations, and anticorrelations, between datasets. They are due not only to the correlations of experimental uncertainties within datasets, but also to the use of a common set of smooth underlying PDFs, and the correlations of the theory uncertainties. The correlations within each process are generally larger than those between processes. This suggests that the combined effects of the correlations due to the use of a common factorization scale and the correlations induced by the smoothness of the PDFs is small compared to the correlations from the renormalization scale.

In Fig.~\ref{fig:Pdiag} we show the percentage uncertainties of the autopredictions: $\sqrt{P_{ii}}$, compared to the purely PDF uncertainties $\sqrt{X_{ii}}$ to aid comparison. The correlated autoprediction uncertainties are generally of similar size to the PDF uncertainties; they are 
rather larger for some of the DY datasets and JETS, but are actually smaller for most of the DIS NC data (most remarkably for the HERA data at small $x$), and some DY data. So the full autopredictions are not only more accurate: they are also more precise. This increase in precision must be taken with a pinch of salt, since it depends to some extent on the assumptions made in modelling the prior MHOU~\cite{AbdulKhalek:2019ihb}, in particular the choice of independent scales, and the scheme through which they are combined into the theory covariance matrix $S$. In particular the aggressive reduction in the small $x$ uncertainties for the HERA NC autopredictions seen in Fig.~\ref{fig:Pdiag} may be due to the adoption of the same factorization scale for both singlet and nonsinglet, which overconstrains the singlet evolution at small $x$ \cite{Harland-Lang:2018bxd}. We leave the relaxation of these kinds of assumptions for future work.

As expected all of the autopredictions uncertainties are smaller than would be obtained by the standard prescription of adding PDF uncertainties and theory uncertainties in quadrature~\cite{AbdulKhalek:2019ihb}, which ignores both learning and correlation. However the conservative approach overestimates the correlated uncertainty for almost all datapoints, typically by a factor of two or more, particularly those for which the theoretical uncertainty is larger than the PDF uncertainty. The only data for which the conservative prescription works well are ratio data (for example the NMC $d/p$ data), for which theoretical uncertainties are very small.

To understand better how these changes in uncertainty arise, we show in Fig.~\ref{fig:Scpts} the contributions to the diagonal elements of the correlated theory uncertainty (the second term in Eq.~(\ref{eq:PNN}), normalised to the uncorrelated elements $S_{ii}$. The `learning' of the theoretical uncertainty, given by the contribution $-S(C+S)^{-1}S$ (note that $S-S(C+S)^{-1}S$ is equal in the one parameter example described in Sec.~\ref{subsec:autoprediction} to $ZS$) is very significant, reducing the prior uncertainty $S$ almost to zero for NC DIS and DY (where there is considerable data), and by an order of magnitude for DIS CC, JETS and TOP. Probably more flexibility is required in the modelling of the prior for this uncertainty. However the PDF fluctuations $S(C+S)^{-1}X(C+S)^{-1}S$ (note that $S-S(C+S)^{-1}S+S(C+S)^{-1}X(C+S)^{-1}S$ is equal in the one parameter example described in Sec.~\ref{subsec:autoprediction} to $\overline{Z}S$) undo much of this learning, though for all data points this effect is insufficient to take the ratio to $S$ above one. 

A similar breakdown of the contributions to the diagonal elements of the correlated PDF uncertainty (the first term in Eq.~(\ref{eq:PNN}), which can be expanded as in Eq.~(\ref{eq:Xalgebra2})), normalised to the uncorrelated elements $X_{ii}$, is shown in Fig.~\ref{fig:Xcpts}. The correlation terms $-S(C+S)^{-1}X-X(C+S)^{-1}S$ are indeed very large, as anticipated in Ref.\cite{Harland-Lang:2018bxd}, in particular for data with relatively large theoretical uncertainty (such as HERA NC at small $x$, or JETS), sufficient there to overwhelm $X$ and give a negative result. However they can also be positive for some data (such as JETS). In any event, the addition of the PDF fluctuation term $S(C+S)^{-1}X(C+S)^{-1}S$ (remember the  decomposition Eq.~(\ref{eq:Xalgebra2}) of the total correlated PDF uncertainty) is always sufficient to restore positivity of the correlated PDF uncertainty, and can in some situations (where $S$ is large, in particular for JETS, but also some DIS CC and DY data) take the total correlated PDF uncertainty above the uncorrelated result $X$. Thus the correlations, while generally reducing uncertainties, can in some circumstances increase them, in contrast to learning which always reduces them.

For autopredictions we expect high levels of learning and correlation because we are making predictions for exact repeats of experiments already in the fit. As we noted for the shifts however, removing a smaller dataset from the fit has little effect in the PDFs, so we might expect similar effects for genuine predictions of processes already included in the fit, particular if they are in a similar kinematic region. 

\subsection{Predictions for Top}
\label{subsec:topnhiggs}

\begin{figure}[b!]
    \begin{center}
    \makebox{\includegraphics[width=0.46\columnwidth]{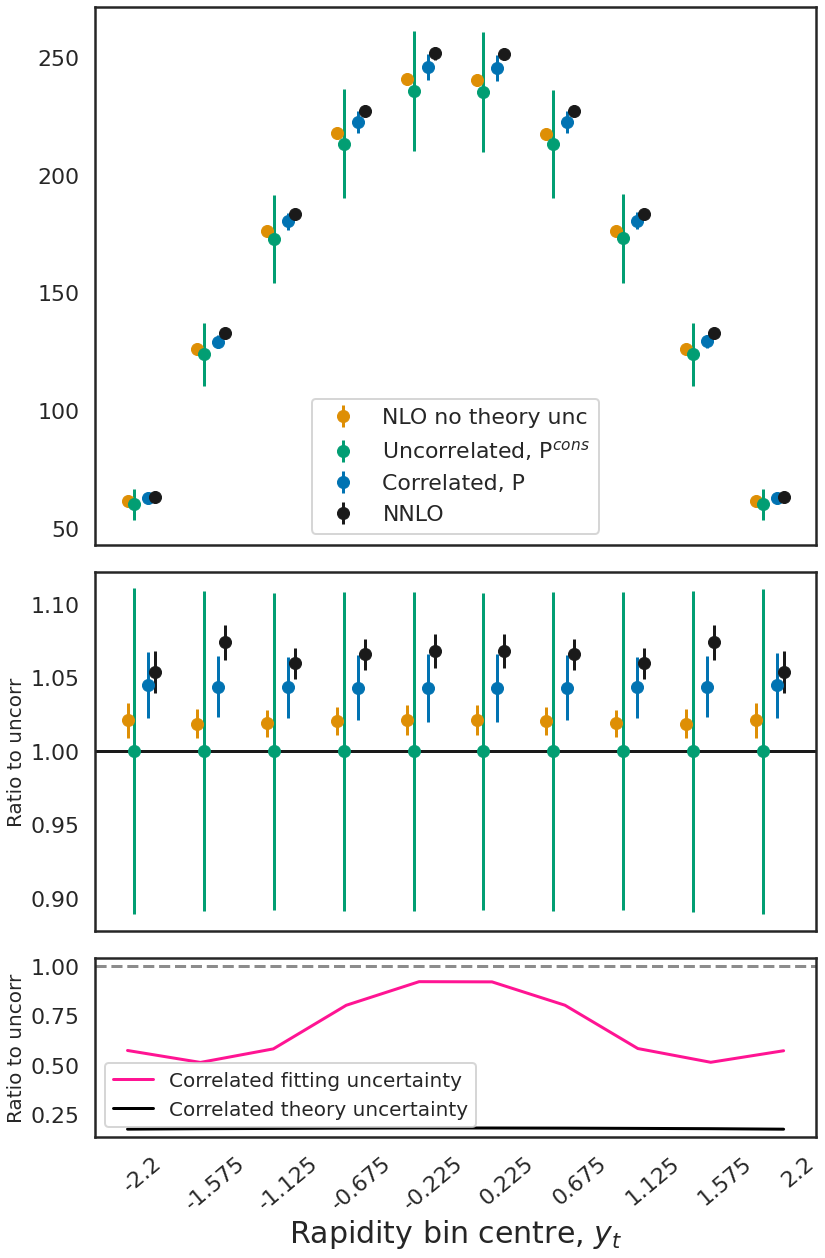}}
   \makebox{\includegraphics[width=0.46\columnwidth]{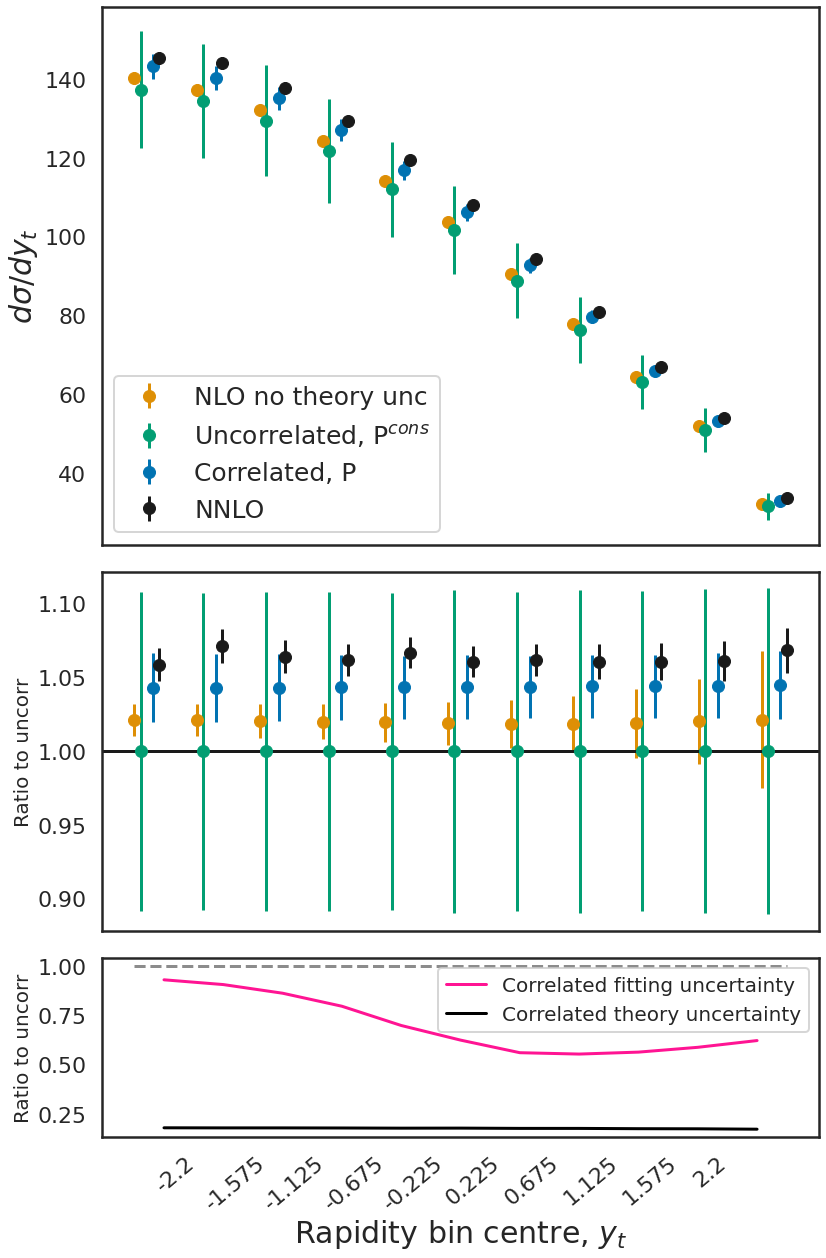}}
    \end{center}
  \vspace{-0.55cm}
  \caption{The upper two panels show predictions for $t\bar{t}$ unnormalized rapidity distribution data taken at 13~TeV by CMS, the dilepton rapidity distribution \cite{Sirunyan:2018ucr} (left) and the lepton+jets distribution \cite{Sirunyan:2018wem} (right). The four predictions show: the NLO fit with no MHOUs, PDF error only; the combined PDF and MHOU fit, ignoring correlations (thus $\sqrt{P_{II}^{\rm con}}$); the correlated result including the shift, and uncertainty computed using the simplified result (thus $\sqrt{P_{II}^{\rm sim}}$); the result with the same shift, but with the correlations included exactly (thus $P_{II}$), and the NNLO result with no MHOU. In the middle panels the same is shown, but normalized to the uncorrelated result. In the lower panels we show the fractional reduction in the PDF uncertainty and the theory uncertainty due to the inclusion of the correlations.}
  \label{fig:CMSttbar}
\end{figure}

We now consider genuine predictions, for experiments not used in the PDF fit. These are of two kinds: those for datasets obtained through processes already contained in the fit, and those for completely new processes. For the former we consider the ${\rm t}\bar{\rm t}$ production rapidity distributions (dilepton and lepton+jets) measured by CMS at 13 TeV \cite{Sirunyan:2018ucr,Sirunyan:2018wem}. We chose these datasets for two reasons: firstly, the MHOU is large compared to the experimental uncertainty; secondly, the fit in Ref.~\cite{AbdulKhalek:2019ihb} contains the total ${\rm t}\bar{\rm t}$ cross-sections at 7, 8 and 13 TeV, and the normalized rapidity distributions at 8 TeV, all from both ATLAS and CMS. Both these factors mean that we expect to see some of the largest possible effects due to correlations between the theoretical uncertainties of the data in the PDFs and the theoretical uncertainties of the 13 TeV rapidity distributions.

To make genuine predictions, including all correlations, we need to first calculate the covariance matrix for the MHOU of the predictions, $\Stil_{IJ}$, and its cross-covariance with the MHOU of the theoretical predictions for the data used in the PDF fit, 
$\Shat_{Ij}$ (the indices $I,J$ running over the predicted data points, while $i,j$ run over the data included in the PDF fit): these are in fact the same as if we were planning to include data for the new process in a PDF fit, since the complete covariance matrix for the MHOU would be then of the form Eq.~(\ref{eq:covmatglobal}). Similarly we need to compute, using the fitted PDF replicas the covariance matrix $\Xtil_{IJ}$ of the PDF uncertainty for the new predictions, and the cross-covariance $\Xhat_{Ij}$ with the PDF uncertainty of the observables used in the fit. All of these matrices are required for an exact calculation of the correlated shifts in the theoretical predictions $\delta\Ttil_I$, Eq.~(\ref{eq:shifttilNN}) and the covariance matrix $\Ptil_{IJ}$ of their combined PDF and correlated theoretical uncertainties  Eq.~(\ref{eq:PtilNN}). 

Predictions for the CMS 13 TeV ${\rm t}\bar{\rm t}$ production rapidity distributions were computed using the same tool chain as in Ref.~\cite{Ball:2017nwa} for the 8 TeV distributions: NLO theoretical predictions were generated with {\tt Sherpa}~\cite{Gleisberg:2008ta}, in a format compliant with {\tt APPLgrid}~\cite{Carli:2010rw}, using the {\tt MCgrid} 
code~\cite{DelDebbio:2013kxa} and the {\tt Rivet}~\cite{Buckley:2010ar} analysis package, with {\tt OpenLoops}~\cite{Cascioli:2011va} for the NLO 
matrix elements. Renormalization and factorization scales have been chosen based on the recommendation of 
Ref.~\cite{Czakon:2016dgf} as $H_T/4$.

The results of these calculations are shown in Fig.~\ref{fig:CMSttbar}. The prior theoretical uncertainty in the original prediction is around 10\%, considerably greater than the PDF uncertainty, as expected since the hard cross-sections are only computed at NLO. The correlated shift is sizeable, around 5\%, and almost fully correlated across all the rapidity distributions. This is because these are unnormalized distributions, and thus have an overall theoretical normalization uncertainty which is strongly correlated to the measurements of the total ${\rm t}\bar{\rm t}$ cross-sections at 7, 8 and 13 TeV by ATLAS and CMS included in the PDF fit. This is confirmed by breaking down the contributions to the shift Eq.~(\ref{eq:shiftpredf}) from the various data points included in the fit: the results are shown in Tab.~\ref{tab:deltilcons}. All of the shift comes from the six total cross-section measurements, while the 8 TeV normalized rapidity distributions push it down again by  around 25\%. The remaining data make almost no contribution. Nevertheless, the shift is still rather less than the theoretical uncertainty in the original prediction, as expected from the shifts in the nuisance parameters for the renormalization scale variation for top processes shown in Fig.~\ref{fig:nuisancephys}. 

\begin{table}[t!]
  \centering
  \scriptsize
  \renewcommand{\arraystretch}{1.4}
  \begin{tabular}{|llll|llll|l|}
   \hline
 {\bf ATLAS} &&&& {\bf CMS} &&&& {\bf Other}\\
 {\it tot} &&&{\it diff} &{\it tot}&&&{\it diff} & \\
  7 TeV &  8 TeV & 13 TeV & 8 TeV &7 TeV & 8 TeV &13 TeV & 8 TeV & \\
    \hline
 0.37 & 0.11 & 0.24 & -0.21 & 0.26 & 0.21 & 0.07 & -0.04 & -0.01 \\
\hline
  \end{tabular}
\caption{The fractional contributions of different data sets included in the fit to the
shifts in the top rapidity distributions, averaged over all 21 data points.}
\label{tab:deltilcons}
\end{table}

\begin{table}[t!]
\centering
\begin{tabular}{|l||cc|c|}
\hline
                   & \textbf{dilepton} & \textbf{lepton+jet} & \textbf{combined} \\
                   \hline
No th uncs         &    0.55         &    0.37         & 0.46             \\
Uncorr th uncs     &   0.57         &    0.42         & 0.49               \\
Correlated th uncs &     0.49        &   0.37        & 0.43         \\
NNLO, no th uncs &    0.45        &    0.39     & 0.42          \\
\hline
\end{tabular}
\caption{The experimental $\chi^2$ per data point for the CMS 13 TeV top dilepton and lepton+jet rapidity distributions, comparing the original result of the NLO fit with no theory uncertainties to the fit with theory uncertainties, and then including the correlated shift in the autopredictions. Also shown for comparison is the result in a NNLO fit with no theory uncertainties, only PDF uncertainties.}
  \label{tab:topchisq}
\end{table}

\begin{figure}[t!]
    \begin{center}
    \makebox{\includegraphics[width=0.4\columnwidth]{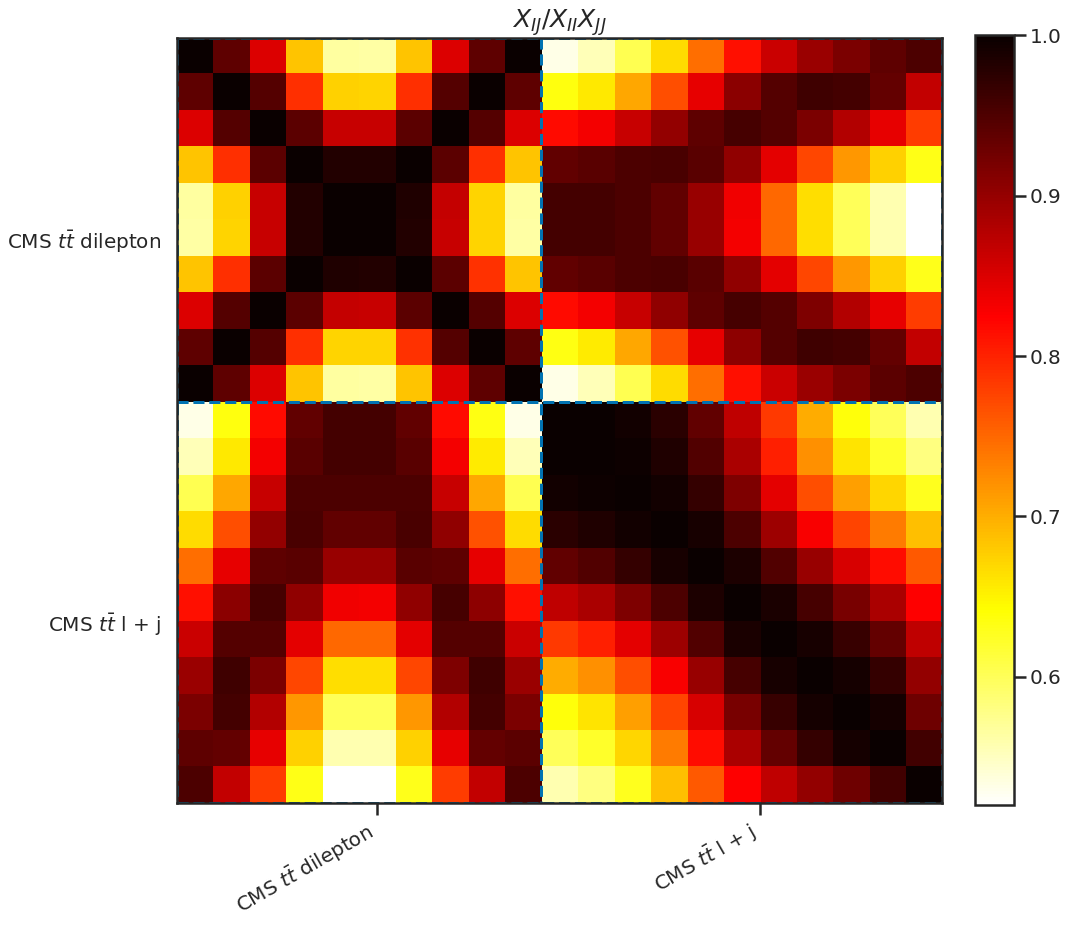}}
   \makebox{\includegraphics[width=0.4\columnwidth]{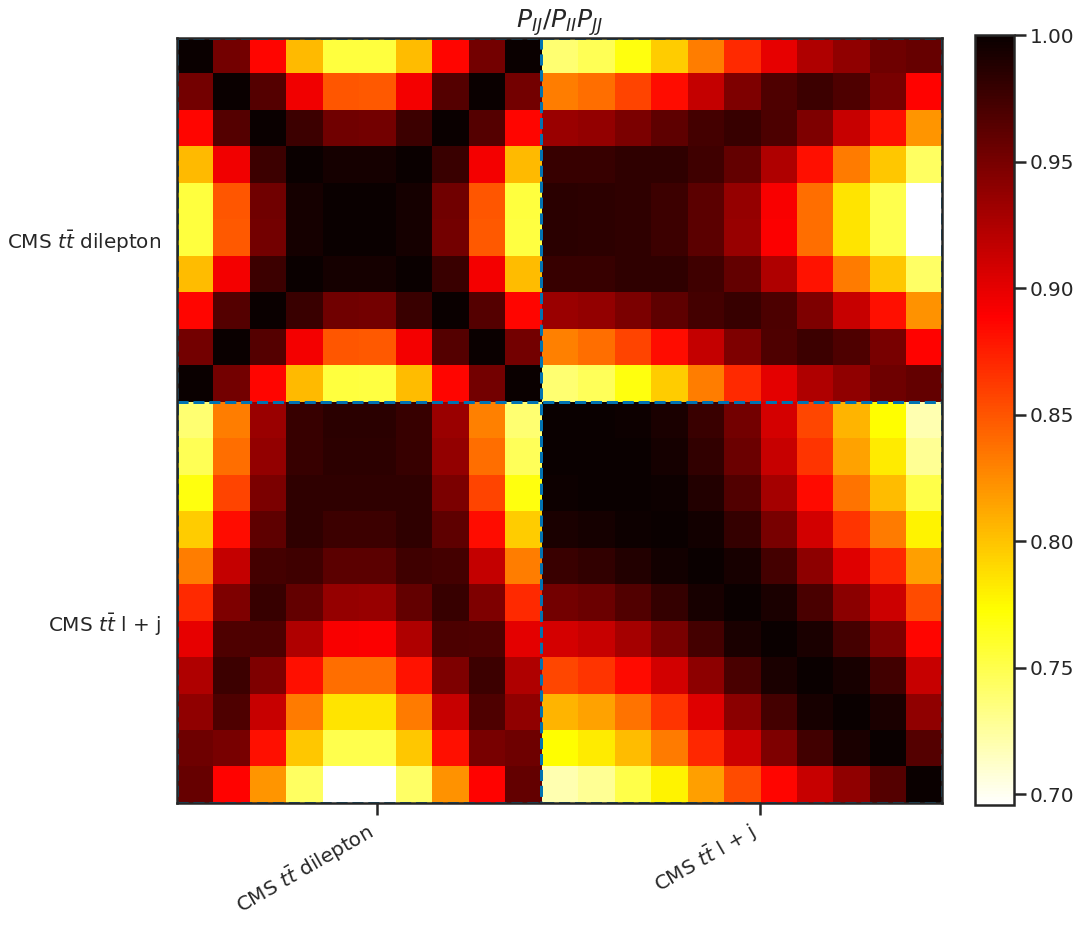}}
    \end{center}
  \vspace{-0.55cm}
  \caption{The left hand plot shows the correlation matrix $\Xtil_{IJ}/\sqrt{\Xtil_{II}\Xtil_{JJ}}$ of the contribution of the PDF uncertainties to the predictions for the 13~TeV rapidity distributions by CMS: the right hand plot shows the correlation matrix $\Ptil_{IJ}/\sqrt{\Ptil_{II}\Ptil_{JJ}}$ of the total uncertainties including the correlated theoretical uncertainties. Note the expanded scales on the heat maps, different in each plot.} 
  \label{fig:CMSttbarcorrlns}
\end{figure}

We can compare the shift due to correlations to that from going from NLO to NNLO. We thus show in Fig.~\ref{fig:CMSttbar} the results of a complete NNLO calculation (without theory uncertainties): the NNLO corrections also increase predictions by 5-8\%.  It is very interesting that the shift, driven by the data for the ${\rm t}\bar{\rm t}$ total cross-sections at 7, 8 and 13 TeV, largely accounts for the NNLO correction: the data know that the NLO theoretical predictions are on the low side, and this information is carried over into the prediction for the 13 TeV rapidity distributions.  Indeed, we compare the experimental $\chi^2$ for these data in the various calculations in Tab.~\ref{tab:topchisq}, while the theory uncertainties increase the $\chi^2$ (due presumably to the other top data being deweighted in the fit), the shift gives a significant improvement both for dileptons and lepton+jets, comparable to that obtained with the complete NNLO corrections. So the shifts provide a new method for using experimental data to make improved theoretical predictions through the learning of theoretical uncertainties. The method should be particularly effective when there is substantial data on the process to be predicted already included in the PDF fit, as is the case here.

The middle panels of Fig.~\ref{fig:CMSttbar} shows the same points as the top panel, but as a ratio to the conservative result, making the uncertainties more visible. Comparing the uncertainties, the difference between the uncorrelated and correlated uncertainty is striking; the correlated uncertainties are much smaller than the uncorrelated. The correlated uncertainties are however still larger than the purely PDF uncertainties, but the very large theoretical uncertainty has been substantially reduced. So not only are the correlated predictions more accurate, they are also more precise.  Despite this significant shrinking of uncertainties, the correlated predictions are still compatible with the NNLO result, thanks to the shift in central values. While the conservative prescription is also compatible with the NNLO result, it is immediately clear from the plot that it is inferior to the correlated prediction.

The breakdown of the reduction in uncertainties due to the correlations is shown in the lower panels of Fig.~\ref{fig:CMSttbar}. The correlated theory uncertainty is substantially reduced (uniformly across the rapidities), due to the learning of the normalization from the data already included in the fit, while the correlated PDF uncertainty is reduced rather less: as much as a factor of two when the differential cross-section is small, but hardly at all when it is large. So here the dominant effect is clearly the learning of the theoretical uncertainty in the overall normalization.

The theoretical uncertainties in the theoretical predictions are strongly correlated amongst themselves, and between the two rapidity distributions: in Fig.~\ref{fig:CMSttbarcorrlns} we show the correlation matrices for the PDF uncertainties, $\Xtil_{IJ}$, and that of the correlated prediction $\Ptil_{IJ}$. We see that while the predictions are more than 50\% correlated across the range of rapidities by the PDF, when the correlated theoretical uncertainties are also included all points are rather more correlated, to more than 70\%. The pattern of correlations reflects the symmetry in the dilepton distribution and asymmetry in the lepton+jet distribution: the least correlated points are those with the greatest rapidity separation.

\subsection{Predictions for Higgs}
\label{subsec:higgs}

\begin{figure}[t!]
    \begin{center}
    \makebox{\includegraphics[width=0.6\columnwidth]{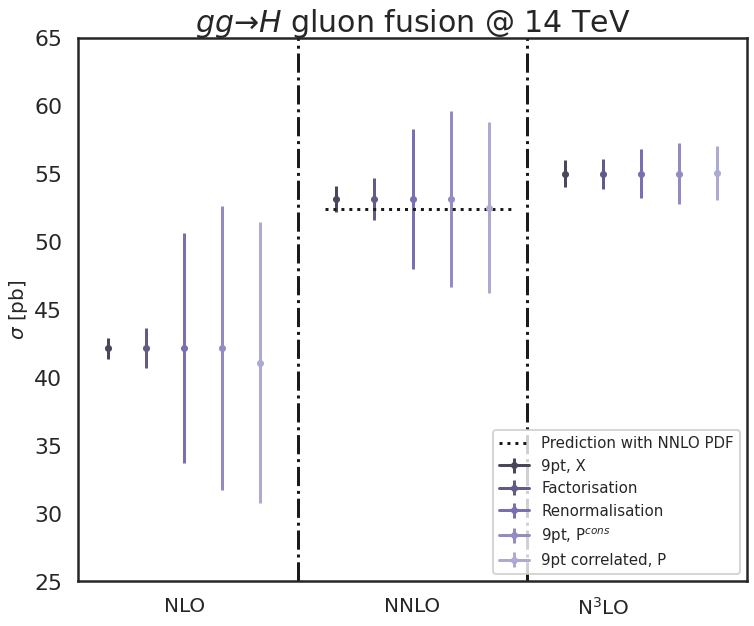}}
    \end{center}
  \vspace{-0.55cm}
  \caption{Predictions for the Higgs total cross-section at 14 TeV, made using a variety of approximations. All results use NLO PDFs, while the Higgs total cross-section is computed at NLO (left panel), NNLO (centre panel) and N3LO (right panel). In each panel, we then have, from left to right: MHOU included only in the PDF determination in the 9pt scheme; the same but with the factorization scale uncertainty (MHOU in PDF evolution) included in quadrature; the same but with instead the renormalization scale uncertainty (MHOU in the Higgs cross-section); the total PDF uncertainty and 9pt MHOU combined in quadrature, as recommended in Ref.~\cite{AbdulKhalek:2019ihb}; the total PDF plus 9pt MHOU, but now including also the shift and the correlation between theoretical uncertainties. In the centre panel we also show the NNLO prediction with NNLO PDFs (but no theoretical uncertainties), as a dashed line. }
  \label{fig:Higgs}
\end{figure}

As an example of a process not included in the PDF fit we consider the prediction for the total cross-section for Higgs production in gluon fusion at 14 TeV. For the calculation of the cross-section we performed calculations using
{\tt  ggHiggs}~\cite{Ball:2013bra,Bonvini:2014jma,Bonvini:2016frm}.
Renormalization and factorization scales are set to $m_h/2$, and the computation is performed using rescaled effective theory.

Our results are shown in in Fig.~\ref{fig:Higgs}. The PDFs are still the NLO PDFs with MHOU from Ref.~\cite{AbdulKhalek:2019ihb}, but the Higgs total cross-sections are computed at NLO, NNLO and N3LO. At NLO the MHOU in the Higgs cross-section, estimated by varying the renormalization scale, completely dominates all other uncertainties (the PDF uncertainty, the theoretical uncertainty in the PDFs, and the scale uncertainty in the PDF evolution, estimated by factorization scale variation), so the effect of correlations in the MHOU is completely negligible. However when the cross-section is computed at NNLO, the renormalization scale uncertainty is rather smaller, and at N3LO it becomes more comparable to the other sources of uncertainty. The shift due to the correlation between MHOUs is always small compared to the overall uncertainty, and becomes smaller still as the perturbative order of the cross-section is increased, as expected. This is because, unlike in the top predictions, data for this process are not included in the fit, so the renormalization scale uncertainty is completely uncorrelated. It is interesting to note however that the small shift due to the correlation in factorization scale takes the NNLO prediction very close to the result computed with NNLO PDFs (though the coincidence is surely accidental). 

Unlike for the top predictions, the effect of the correlation on the size of the overall uncertainty is small. Again, this is because the PDF contains far less information about Higgs production than it does about top production. We saw in Fig.~\ref{fig:P} that the information propagated through to the correlated uncertainties was primarily through the renormalization scales, and that correlation due to the factorization scale was a lot weaker. Higgs production, being a new process, is therefore only impacted through the weak factorization scale correlations, and so the reduction in uncertainties is small. Overall, the uncorrelated conservative prescription~\cite{AbdulKhalek:2019ihb} is comparable in this case to the fully correlated one. Note that if we were to perform these studies with NNLO (or indeed N3LO) PDFs which include MHOU, the MHOU in the PDF determination would have presumably been rather smaller than at NLO, and thus the effect of correlations in the MHOU, in particular the shift, but also the effect on the size of the uncertainty would be even smaller than the small corrections we see here. 

From these examples of autopredictions, and genuine predictions for top and Higgs, we have seen that the extent of the shift and correlation can vary quite significantly, depending on the type of prediction being made and what information is already contained in the PDFs. The conservative prescription recommended in Ref.~\cite{AbdulKhalek:2019ihb} is certainly not appropriate in general, as the full inclusion of correlations can be quite substantially reduce uncertainties, as we saw both for the autopredictions and top predictions. However, when predicting a new process for which the PDF contains little information about correlated theoretical uncertainties, unsurprisingly the impact of correlations is small and the conservative prescription is quite sufficient.

\section{Summary}

In this paper we studied in detail the correlation between theoretical uncertainties in the calculations used in the determination of PDFs in a global fit, as formulated in Ref.~\cite{AbdulKhalek:2019ihb}, and the theoretical uncertainties in the predictions made using these PDFs. We began by recasting the theoretical uncertainties using nuisance parameters, determined replica by replica, which carry all the information about the effect of the experimental data on the theoretical uncertainties. Using increasingly realistic models of the fitting procedure, we produced analytical formulae for computing fully correlated predictions. In the process we identified three distinct but related effects, each of which has a significant impact on the final theoretical predictions:
\begin{itemize}
\item {\bf Shifts in central values.} These are an effect of Bayesian learning: just as we can use experimental data to determine PDFs, so we can also use it to identify theoretical corrections that improve the agreement between data and theory, while remaining within theoretical uncertainties. The correlations between theoretical uncertainties in the fit and those in predictions then lead to more accurate predictions. This effect was first identified in Sec.~\ref{subsec:puretheory}.
\item {\bf Learning of  theoretical uncertainties.} A second consequence of Bayesian learning is a reduction in theoretical uncertainty, due to the information provided by the data in the PDF fit, which through correlation of theoretical uncertainties can lead to a corresponding reduction in the theoretical uncertainties in predictions. This effect was also identified in Sec.~\ref{subsec:puretheory}, and is complementary to the shift in central values.
\item {\bf Correlation in theoretical uncertainties.} The third effect is that the correlation between the theoretical uncertainties in the fit and the theoretical uncertainties in the predictions lead to a change in the PDF uncertainties in the prediction, even in situations where there is no shift, thus avoiding any `double counting' of the theoretical uncertainty. The existence of this effect was first noted in Ref.\cite{Harland-Lang:2018bxd}, and identified as an effect distinct from Bayesian learning in Sec.~\ref{subsec:phenomenology}.
\end{itemize}

While these three effects were first identified in the simple models of Sec.~\ref{sec:generic}, we showed that they are all present in the one parameter fits of Sec.~\ref{sec:oneparameter} and the more realistic fits with multiple parameters in Sec.~\ref{sec:corrlnpdffits}. Using the NNPDF3.1 NLO global fits with MHOU~\cite{AbdulKhalek:2019ihb}, we demonstrated in Sec.~\ref{sec:numeric} that the shifts can give sensible estimates of NNLO corrections, and thereby reduce the $\chi^2$ to the experimental data. We also showed that while the uncertainty in NLO predictions is still a sum in quadrature of the theoretical uncertainty and the PDF  uncertainty (which also includes a theory uncertainty), this sum can be significantly reduced, depending on the relative size of the theoretical and experimental uncertainties. Consequently the `conservative' prescription of Ref.~\cite{AbdulKhalek:2019ihb}, where the theory uncertainty in the prediction is combined in quadrature with the PDF uncertainty, is indeed conservative. We expect these conclusions to also hold in global PDF fits with fixed parametrization and tolerance \cite{Bailey:2020ooq,Hou:2019efy}, if these were to include MHOU in the PDF fit. 

The degree of correlation is highly dependent on the type of prediction being made. For the autopredictions (predictions for new measurements of the same data points as those included in the fit), Sec.~\ref{subsec:autopredictions}, where there is maximal correspondence between the data in the fit and the predictions being made, the correlation is very high, leading to shifts that improve the quality of the fit to the data, together with a significant reduction in uncertainties, in some cases down to a small fraction of the uncorrelated values. For genuine predictions for new measurements of processes already included in the PDF fit, such as the new measurements of differential top production discussed in Sec.~\ref{subsec:topnhiggs}, we observe that the shift takes the correlated NLO predictions very close to the NNLO prediction, with a significant reduction in uncertainties: the prediction is both more accurate and more precise. For Higgs production, discussed in Sec.~\ref{subsec:higgs}, a process not included in the PDF  fit, the level of correlation is much smaller, since the dominant uncertainty (the MHOU in the hard cross-section) is uncorrelated with the MHOU of the fitted processes. In this case the shift is well within uncertainties, and the reduction in uncertainty very modest, so here the use of the conservative prescription Ref.~\cite{AbdulKhalek:2019ihb} is entirely appropriate. We expect this to be true of predictions for any new process.

Thus our main conclusion is that when using PDFs which include MHOUs, taking account of the correlations between the MHOU included in the determination of the PDFs and the MHOU in the prediction can result in a significant improvement in both accuracy and precision. This is especially true in the case where the predicted process is among those included in the fit. However the correlated predictions must be treated with care, since their reliability relies to some extent on the generality of the prior estimation of the MHOU: if unjustified assumptions are made in the choice of prior, the uncertainty estimates in the correlated predictions may be too aggressive. For these reasons the conservative prescription, as an upper bound on the overall uncertainty, may sometimes be preferable, especially for predictions of new processes. 

In order to calculate fully correlated predictions and uncertainties, one requires besides the PDF replicas some additional  information: the cross-correlations between the theoretical uncertainties in the prediction and those in the theoretical calculations used to determine the PDFs,  $\Shat_{Ij}$; and the cross-correlations between the PDF uncertainties in the prediction and all the calculations included in the fit, $\Xhat_{Ij}$.  In the future, it may be possible to present this information in separate NNPDF deliverables to facilitate the calculation of the correlation effects.

Although we presented our numerical study of correlations in the context of MHOUs, we would expect similar results for other kinds of theoretical uncertainty, such as nuclear uncertainties, higher twist uncertainties, or indeed parametric uncertainties: once the theory covariance matrix has been computed, the linear algebra has no concern for the type of theoretical uncertainty it contains. This suggests a new technique for determining external parameters in PDF fits, such as quark masses or electroweak parameters, taking full account of all correlations with the PDFs and theoretical uncertainties. We hope to explore this possibility in the near future.

\paragraph{Acknowledgments.}
We would like to offer our warmest thanks to Emanuele Nocera for his assistance in the numerical computations. We would also like to thank an anonymous referee for comments which significantly improved the final draft of the paper. R.D.B. and R.L.P. are supported by the UK Science and Technology Facility Council through grants ST/P000630/1 and ST/R504737/1.
\bibliography{corrthunc}

\end{document}